\documentclass[useAMS,usenatbib]{mn2e}

\usepackage{hyperref}
\usepackage{graphicx}
\usepackage{times}
\usepackage{natbib}
\usepackage{amssymb,amsfonts,txfonts}
\usepackage[latin1]{inputenc}

\newif\ifAMStwofonts
\AMStwofontstrue

\newcommand{\mincir}{\raise
  -2.truept\hbox{\rlap{\hbox{$\sim$}}\raise5.truept \hbox{$<$}\ }}
\newcommand{\magcir}{\raise
  -2.truept\hbox{\rlap{\hbox{$\sim$}}\raise5.truept \hbox{$>$}\ }}
\newcommand{\siml}{\raise
  -2.truept\hbox{\rlap{\hbox{$\sim$}}\raise5.truept \hbox{$<$}\ }}
\newcommand{\simg}{\raise
  -2.truept\hbox{\rlap{\hbox{$\sim$}}\raise5.truept \hbox{$>$}\ }}

\renewcommand{\vec}[1]{\bmath{#1}}
\newcommand{\be}{\begin{equation}}
\newcommand{\ee}{\end{equation}}
\newcommand{\ba}{\begin{eqnarray}}
\newcommand{\ea}{\end{eqnarray}}
\newcommand{\brr}{\begin{array}}
\newcommand{\err}{\end{array}}
\newcommand{\bc}{\begin{center}}
\newcommand{\ec}{\end{center}}

\newcommand{\fl}{\,{\rm erg\,s^{-1}cm^{-2}}}

\newcommand{\omegam}{$\Omega_{m}$}
\newcommand{\omegab}{$\Omega_{b}$}
\newcommand{\omegade}{$\Omega_{e,de}$}
\newcommand{\lcdm}{$\Lambda$CDM}
\newcommand{\MpcI}{\,\rm{Mpc}^{-1}}

\newcommand{\chandra}{\textit{Chandra}}
\newcommand{\referee}[1]{#1}
\newcommand{\refereebis}[1]{#1}
\newcommand{\refereetris}[1]{#1}
\newcommand{\numero}{}

\begin{document}
\title[Dark energy with galaxy clusters]
  {Probing dark energy with the next generation X-ray surveys of galaxy 
clusters} 
\author[Sartoris et al.]{B.\,Sartoris$^{1,2,3}$, S.\,Borgani$^{1,3,4}$,
  P.\,Rosati$^2$ \& J.\,Weller$^{5,6,7}$\\~\\
$^1$ Dipartimento di Fisica, Sezione di Astronomia, Universit\`a di
Trieste, Via Tiepolo 11, I-34143 Trieste, Italy\\ \texttt{e-mail:
sartoris@oats.inaf.it}\\ 
$^2$ ESO-European Southern Observatory, D-85748 Garching bei
M\"unchen, Germany\\ 
$^3$ INAF-Osservatorio Astronomico di Trieste, Via Tiepolo 11, I-34143
Trieste, Italy\\ 
$^4$ INFN, Sezione di Trieste, Via Valerio 2, I-34127 Trieste, Italy\\
$^5$ University Observatory, Ludwig-Maximillians University Munich,
Scheinerstr. 1, 81679 Munich, Germany\\
$^{6}$ Excellence Cluster Universe, Boltzmannstr. 2, 85748 Garching, Germany\\
$^{7}$ Max-Planck-Institut f\"{u}r extraterrestrische Physik, Giessenbachstrasse, 85748 Garching, Germany
}
\maketitle

\begin{abstract}

 We present forecasts on the capability of future wide-area high-sensitivity
  X-ray surveys of galaxy clusters to yield constraints on the
  parameters defining the Dark Energy (DE) equation of state (EoS). Our
  analysis is carried out for future X-ray surveys which have enough
  sensitivity to provide accurate measurements of X-ray mass proxies and Fe-line based redshifts
  for about $2\times 10^4$ \numero clusters. We base our analysis on
  the Fisher Matrix formalism, by combining information on the cluster
  number counts and power spectrum, also including, for the first time 
  in the analysis of the large scale cluster distribution, the effect of
  linear redshift space distortions. This study is performed with
  the main purpose of dissecting the cosmological information provided
  by geometrical and growth tests, which are both included in the
  analysis of number counts and clustering of galaxy clusters.  We
  compare cosmological constraints obtained by assuming different
  levels of prior knowledge of the parameters which define the
  relation between cluster mass and X-ray observables. This
  comparison further demonstrates the fundamental importance of having
  a well calibrated observable-mass relation and, most importantly,
  its redshift evolution. Such a calibration can be achieved only by
  having at least $\sim 10^3$ \numero net photon counts for each
  cluster included in the survey, with sufficient angular resolution.
  We show that redshift space distortions in the
  power spectrum analysis carry important
  cosmological information also when traced with galaxy clusters. We
  find that the DE FoM increases by a factor of 8 \numero when
  including the effect of such distortions. Besides confirming the
  potential that large cluster surveys have in constraining the nature
  of Dark Energy, our analysis emphasizes that a full
  exploitation of the cosmological information carried by such surveys
  requires not only a large statistic but also a robust measurement of
  the mass proxies and redshift for a significant fraction of the cluster sample, which 
  ought to be derived from the {\sl same} X-ray survey data. 
  This will be possible with future X-ray surveys, such as those envisioned with the Wide Field X-ray Telescope, with
  an adequate combination of survey area, sensitivity and angular
  resolution.
 
\end{abstract}

\section{Introduction}
\label{s:intro}

A number of independent cosmological observations, ranging from
type-Ia supernovae (SNIa) \citep[e.g.][]{riess07,perlmutter99}, to Cosmic Microwave
Background (CMB) anisotropies \citep[e.g.][]{komatsu11,larson11} and
Large Scale Structure (LSS) \citep[e.g.][]{vikhlinin09c,reid10},
convincingly show that the Universe is undergoing a phase of
accelerated expansion. One of the main challenges of modern cosmology
is in fact understanding the source of such acceleration. To this
purpose, a number of models have been proposed that modify the two
pillars of modern cosmology, general relativity and the standard
model of fundamental interactions \citep[e.g.][and references
therein]{silvestri09}. Models that modify the latter and, therefore,
the energy-momentum tensor in the Einstein equation are, for example,
the scalar field models like quintessence
\citep[e.g.][]{caldwell09,doran03,zlatev99,ratra88}, k-essence
\citep[e.g.][]{tsujikawa10,mukohyama04}, coupled Dark Energy
\citep[e.g.][]{amendola00} and Chapligin gas
\citep[e.g.][]{bento02}. Models that modify general relativity can
produce the cosmic acceleration without including a Dark Energy
(DE) component, but they should also satisfy stringent constraint from
local (e.g. solar system) tests of gravity.  Examples of such models
are the braneworld models, like DGP \citep[e.g.][]{dgp,movahed09},
$f(R)$ theories \citep[e.g.][]{hu_sawicki07,brax08,appleby10,sotiriou10} and
scalar-tensor theories \citep[e.g.][]{skordis09}.

The first model proposed to explain cosmic acceleration was based on
the introduction of a cosmological constant $\Lambda$ that can be
thought as a fluid with negative pressure and equation of state (EoS)
which is constant in space and time, $p=w\rho c^2$ with $ w =-
1$. However, a cosmological constant able to drive the accelerated
expansion leads to the well known problem that $\Lambda$ needs to be
so tiny with respect to any natural energy scale, that there is no
theoretical justification for it. Thus, a plethora of models,
characterized by different parametrizations of the DE EoS evolution,
have been proposed \citep[e.g.][]{wetterich04}. In principle, it is
possible to distinguish among different DE models by combining
different cosmological probes, both based on the geometry of the
universal background and the growth of density perturbations (e.g.,
\citealt{albrecht06}).

In this context, clusters of galaxies have long been recognized as
potentially powerful probes of the nature of DE and cosmological
models in general \citep[e.g.][and references therein]{allen11,lombriser10,manera06}.
Clusters of galaxies provide cosmological information in a number of
different ways. The evolution of the cluster space density depends on
cosmological parameters through both the linear growth rate of density
perturbations and  the redshift dependence of the volume
element. The large-scale clustering of galaxy clusters is
also sensitive to cosmological parameters, through the growth rate of
perturbations, which affects both the bias parameter and the
redshift-space distortions, as well as by sampling the shape of the
underlying Dark Matter (DM) power spectrum over a broad range of
wavenumbers. As of today, relatively small samples of $\sim 100$
X-ray selected clusters, originally identified out to $z\simeq 0.8$
by the ROSAT satellite and then followed-up by \chandra\ to obtain
robust mass estimates, have provided interesting DE constraints
\citep[e.g.,][]{vikhlinin09c,mantz10a}, which  complement and
agree with those from CMB and SNIa observations \cite[see
also][]{rapetti05}.  More recently, by using a complete sample
of $\mincir 1000$ nearby clusters at $z\mincir 0.2$, identified in
the ROSAT All-Sky Survey \citep[RASS;][]{truemper93},  the
large-scale power spectrum has been constructed over a fairly wide scale
range \citep[e.g.,][]{antolinez11}.

While the first Sunyaev-Zeldovich (SZ) surveys have now started
producing cluster samples of similar sizes,
\citep[e.g.,][]{ACT11,SPT11,PlanckESZ_sample}, the next generation of
X-ray (e.g.,
eROSITA\footnote{http://www.mpe.mpg.de/heg/www/Projects/erosita/index.php},
WFXT\footnote{http://www.wfxt.eu/}) and optical (e.g.,
DES\footnote{http://www.darkenergysurvey.org/},
EUCLID\footnote{http://sci.esa.int/science-e/www/area/index.cfm?fareaid=102})
surveys are expected to increase by orders of magnitude the number of
galaxy clusters, further extending the redshift range over which they
trace the growth of cosmic structures. Such large cluster surveys have
the potential of placing very tight constraints on different classes
of DE models, possibly finding signatures of departures from the
standard $\Lambda$CDM predictions. Several studies have dealt with
constraints of DE models from future cluster surveys focusing on the
impact of uncertainties in cluster mass estimates
\citep[e.g.][]{battye03,majumdar04,lima05,lima07,cunha09,cunha10,
  basilakos10}. All these analyses generally assume that, when a
cluster is detected and included in a survey, the observable (i.e.,
X-ray luminosity, optical richness) on which the detection is based can be
related to the actual cluster mass through a suitable relation, whose
functional form is assumed to be known and depends on a set of
additional parameters.

A more conservative approach would instead require that for all
clusters included in a survey detailed follow-up observations are
carried out to calibrate suitable and robust mass proxies. Examples of
such mass proxies in X-ray surveys include the total gas mass
\citep[e.g.][]{mantz09II} or the product of gas mass and temperature,
the so-called $Y_X$ parameter \citep{kravtsov06}, which can be
computed when a relatively large number of photons is
available. Clearly, while measuring flux for an X-ray extended source
requires only $\sim 50$ photons, or less for missions with low
background, the measurement of robust mass proxies requires at least $\sim 10^3$
photons.

In this paper, we will derive forecasts for constraints on the EoS of
DE models from future X-ray surveys. We envision that these surveys
will be carried out with a telescope with high-enough sensitivity to
readily provide robust measurements of mass proxies and Fe-line based redshifts for $\magcir 10^4$
clusters, which are all characterized by ``\chandra-quality'' data,
thus avoiding the need of external and time-consuming follow-up
observations. For example, the Wide Field X-ray Telescope (WFXT)
concept was developed to meet such a requirement. This telescope
combines large collecting area with a large field-of-view and sharp
point spread function (PSF) approximately constant over the entire
field of view \citep[e.g.,][]{giacconi09,murray10,rosati10}.

We adopt the
specifications of the WFXT surveys, as an example of next generation X-ray cluster surveys,  and we compute cosmological forecasts
using the well-established Fisher Matrix approach
\citep[e.g.,][]{dodelson03}, to combine information from cluster
number counts and large-scale clustering. We will quantify the
constraints expected on DE models and their dependence on the
knowledge of the relation between X-ray observables and cluster mass,
for a range of survey strategies (i.e. depth vs. sky coverage). In the
course of this study, we will discuss how number counts and the power
spectrum of the large-scale distribution of clusters convey
cosmological information.

As a novel contribution in this paper, we will show how the detection
of Baryonic Acoustic Oscillations (BAOs) and redshift space
distortions (RSDs) on cluster scales can significantly contribute to
constrain cosmological parameters, similarly to a number of previous
studies based on the large scale distribution of galaxies
\citep[e.g.,][]{guzzo08,rassat08,stril10,wang10}.



This paper is structured as follows. In Section 2, we briefly describe
the parametrizations of the DE EoS that we use, and describe
our Fisher-Matrix approach for cluster number counts and power
spectrum.  In Section 3, we first describe the characteristics of the
cluster survey, and derive the forecasts on the constraints on DE EoS
parameters. In this section, we will also quantify the impact that
uncertainties in the scaling relation between X-ray observables and
cluster mass have on such constraints.  We discuss our results and present
our conclusions in Section 4.

\section{Clusters as Dark Energy probes}
\label{s:clus}

The relevant parameters of our analysis are the power spectrum
normalization, $\sigma_8$, the matter density parameter, $\Omega_m$,
and the specific parameters defining the DE EoS.

The reference analysis is
carried out for the standard parametrization of the DE EoS, originally
proposed by \cite{linder03},
\begin{equation}
  w(a)\,=\,w_0\,+\,w_a(1-a)\,,
\label{eq:eos}
\end{equation}
where $a$ is the cosmic expansion factor \citep[see also][]{chevallier01}. This parametrization has
been used in the Dark Energy Task Force reports
\citep[DETF;][]{albrecht06,albrecht09} to assess the constraining
power of different cosmological experiments.

\citet{albrecht06} presented forecasts on the constraints on the $w_0$
and $w_a$ parameters from redshift number counts of cluster surveys.
\citet{mantz10a} derived constraints on these parameters from the
observed evolution of the cluster X-ray luminosity function, using a
combination of nearby clusters selected from the RASS
\citep{truemper93,ebeling98,boehringer04} and medium-distant
clusters selected from the (RASS based) MACS survey
\citep[e.g.][]{ebeling10}. An update on the constraints available at
present on these parameters has been presented by \citet{komatsu11} using
a combination of the 7-year WMAP CMB data, SN-Ia, Big-Bang
Nucleosynthesis results and BAOs traced by the large-scale galaxy
distribution.

Furthermore, we will also assess the constraining power of X-ray
cluster surveys for the class of quintessence models, called Early
Dark Energy (EDE) \citep{wetterich04}. In these models, DE drives not only the
accelerated expansion of the Universe at relatively low
redshift, but also provides a non-negligible contribution at early
times, i.e. before the last scattering surface \citep{doran03}.  A
parametrization of a class of EDE models has been proposed by
\cite{wetterich04} as a function of the amount of DE at $z=0$, the
present EoS parameter, $w_0$, and an average value of the
energy density parameter at early times, \omegade.  The EoS
parametrization that we choose is the one studied by
\citet{grossi09c}:
\begin{equation}
 w\left(z\right) \, = \, \frac{w_0}{(1+C\,\ln(1+z))^2}\,.
\label{eq:eos_ede}
\end{equation}
In the above relation the quantity $C$ is given by
\begin{equation}
 C \, = \, \frac{3\,w_0}
  { \ln\left(\frac{1-\Omega_{e,de}}{\Omega_{e,de}}\right) + 
  \ln\left(\frac{1-\Omega_{m,0}}{\Omega_{m,0}}\right) }\,,
\end{equation}
and characterizes the redshift at which a constant EoS turns into a
different behaviour according to the presence of DE at early times.
Since both EDE and \lcdm\ models have to reproduce the observed
cluster abundance at low redshifts, in the EDE model we expect structures
to form earlier and to have slower evolution of the halo population
that in the \lcdm\ one.

\citet{alam11} used the EDE parametrization by \citet{corasaniti03} to
forecast constraints on these models from the abundance of X-ray
clusters expected in the eROSITA survey \citep[e.g.][]{predehl07} and
from the SZ power spectrum from the South Pole Telescope
\citep[SPT,][]{staniszewski09}. In our analysis, we use a different
parametrization of EDE models, we derive forecasts for high-sensitivity
X-ray surveys, and, particularly, we include the constraints from the
redshift space power spectrum of clusters.  \newline

\subsection{Cluster number counts}
\label{s:fmnc}

The information Fisher Matrix (FM) is defined as
\begin{equation}
 F_{\alpha \beta} \equiv  - \left \langle \frac {\partial^2 \ln  {\cal L}}
{\partial p_{\alpha} \partial p_{\beta}} \right \rangle \, ,
\end{equation}
where ${\cal L }$ is the likelihood of the observable
\citep[e.g.][]{dodelson03}. This can be used to understand how
accurately we can estimate the value of a vector of parameters
$\textbf{p}$ for a given model from one or more data sets, under the
assumption that all parameters have a Gaussian distribution.

Following the approach of \citet{holder01} and \citet{majumdar03},
the FM for the number of clusters, $N_{l,m}$, within the
$l$-th redshift bin and $m$-th bin in observed mass
$M^{ob}$, can be written as
\begin{equation}
  F^N_{\alpha \beta}= \sum_{l,m} \frac{\partial N_{l,m}}{\partial
    p_\alpha}\frac{\partial N_{l,m}}{\partial p_\beta}
  \frac{1}{N_{l,m}}\,, 
\label{eq:fm_nc}
\end{equation}
where the sums over $l$ and $m$ run over redshift and mass intervals,
respectively. In this notation, $M^{ob}_{l,m=0}=M_{\rm thr}(z)$,
where $M_{\rm thr}(z)$ is defined as the threshold value of the
observed mass for a cluster to be included in the survey.
We write the number of clusters expected in a survey with a sky
coverage $\Delta\Omega$, with observed mass between $M^{ob}_{l,m}$ and
$M^{ob}_{l,m+1}$, and observed redshift between $z_l^{ob}$ and $z_{l+1}^{ob}$ as
\begin{eqnarray}
N_{l,m} & = & \Delta\Omega \int_{z_l^{ob}}^{z_{l+1}^{ob}}dz\, 
{dV\over dz d\Omega} \nonumber \\ 
& & \int_{M^{ob}_{l,m}}^{M^{ob}_{l,m+1}} \frac{dM^{ob}}{M^{ob}}  \int_0^\infty dM
\,n(M,z)\,p(M^{ob}\|M)\,.
\label{eq:nln}
\end{eqnarray}
In the equation above, $dV/dz$ is the cosmology-dependent comoving
volume element per unity redshift interval and solid angle, and
$n(M,z)$ the halo mass function. We assume the expression provided by
\citet{jenkins01} for the mass function of halos, with mass computed
at the virial overdensity, 
$\Delta_{\rm vir}=324$,
for the cosmological model assumed in their simulations (see their
equation (B4)). We verified in test cases that all our results are left
unchanged if we use instead the more recent calibration of the mass
function proposed by \citet{tinker08}.  Furthermore, $p(M^{ob}\|M)$ is
the probability to assign to each cluster with true mass $M$ an
observed mass $M^{ob}$. This probability is defined using the
prescription of \citet{lima05}, which takes into account the presence
of a lognormal-distributed intrinsic scatter in the scaling relation
between observable and mass \cite[see also][S10
herefater]{sartoris10}:
\begin{equation}
p(M^{ob}\|M)\,=\,{\exp[-x^2(M^{ob})]\over \sqrt{\left( 2\pi \sigma^2_{\ln
M}\right) }}\,,
\label{eq:prob}
\end{equation}
where
\begin{equation}
x(M^{ob})\, =\, \frac{\ln M^{ob}-B_M-\ln M}{\sqrt{\left( 2 \sigma^2_{\ln
M}\right) }}\,.
\label{eq:m_mo}
\end{equation}
Here $B_M$ is the fractional value of the systematic bias in the mass
estimate and $\sigma_{\ln M}$ is the intrinsic scatter in the
relation between the true and the observed mass. A negative value for
$B_M$ corresponds to a mass underestimate and, therefore, to a smaller
number of clusters included in a survey, for a fixed selection
function. The intrinsic scatter has the effect of increasing the
number of clusters included in the survey. In fact, the number of
low-mass clusters that are up-scattered above the survey mass limit is
always larger than the number of rarer high-mass clusters which are
down-scattered below the same mass limit \citep[e.g.,][ and references
therein]{cunha09}.

Including equation (\ref{eq:prob}) into equation (\ref{eq:nln}), we obtain
\begin{eqnarray}
N_{l,m}& = & \frac{\Delta\Omega}{2} \int_{z_l^{ob}}^{z_{l+1}^{ob}}dz\, 
{dV\over dz d\Omega} \nonumber \\ 
& & \int_0^\infty dM\, n(M,z) \left[{\rm erfc}(x_m)-{\rm erfc}(x_{m+1})
\right]\,
\label{eq:nln2}
\end{eqnarray}
where ${\rm erfc}(x)$ is the complementary error function.

In equation (\ref{eq:nln2}), we assume the sky coverage $\Delta \Omega$ to be
independent of the limiting mass threshold or, equivalently, of the
cluster flux. This formalism can be easily generalized to include a
flux-dependent sky coverage, which is due to the general degradation
of resolution and sensitivity over the the field of view of X-ray
telescopes. Finally, we assume that errors on the cluster redshift
measurements can be ignored (see discussion in Section \ref{s:sur};
see also \citealt{lima07} for a presentation of a method to include
the effect of redshift errors in the computation of the Fisher Matrix
for cluster number counts).

\subsection{Power spectrum}
\label{s:fmps}

A new aspect in this analysis,
with respect to one presented in S10, is the inclusion of the
distorted anisotropic mapping between the real space density field and
measurements in redshift space caused by peculiar velocity
(redshift space distortions). Following \citet{kaiser87}, the
redshift space matter power spectrum $\tilde{P}\left(k,\mu\right)$ in
the linear regime acquires a dependence on the cosine of the angle
between the wave number \textbf{k} and the line-of-sight direction,
$\mu$, according to
\begin{equation}
\tilde{P}\left(k,\mu,z\right) \, = \, \left( b_{eff} + f \mu^2\right)^2
D^2(z)\, P\left(k\right)\,.
\label{eq:kai}
\end{equation}
Here, $P(k)$ is the matter power spectrum in real space, $b_{eff}$ is
the linear bias weighted by the mass function (see equation (20) in S10), and
$f(a) =d{\ln D(a)}/d{\ln a}$ is the so-called growth function,
i.e. the logarithmic derivative of the linear growth rate of density
perturbations, $D(a)$, with respect to the expansion factor $a$. Here
we assume for the bias of halos of mass $M$ the expression provided by
\cite{sheth99}.

The average cluster power spectrum calculated within a given redshift
interval, $\bar{P}_{cl}$, can then be written as
\begin{equation}
\bar{P}^{cl}_{l,m,i}(\mu,k,z_l) = \frac{\int^{z_{l+1}}_{z_l} dz
  \,\frac{dV}{dz} \,\tilde{n}^2(z)\, \tilde{P}(\mu,k,z)}{\int^{z_{l+1}}_{z_l} dz
  \,\frac{dV}{dz}\, \tilde{n}^2(z) }\,,
\label{eq:barpk}
\end{equation}
where $\tilde{n} = \int_0^\infty dM\, n(M,z) \left[{\rm erfc}(x_m)-{\rm
erfc}(x_{m+1})
\right]$.
Therefore, the Fisher Matrix for the power spectrum of galaxy clusters is
\begin{equation}
F_{\alpha \beta}={1\over 8 \pi^2}\,\sum_{l,m,i} 
{\partial \ln{\bar{P}_{cl}(\mu_i,k_m,z_l)}\over \partial p_\alpha} 
{\partial \ln{\bar{P}_{cl}(\mu_i,k_m,z_l)}\over \partial p_\beta}\,
V^{eff}_{l,m,i}k_m^2\Delta k \Delta \mu,
\label{eq:fm_pk}
\end{equation}
\citep[e.g.][]{tegmark97,feldman94,abramo11}
where the sums over $l$, $m$, $i$ run over bins in redshift, wavenumber 
$k$ and angle $\mu$, respectively. 
\referee{The quantity $V^{\mathrm{eff}}(\mu,k,z)$ 
in equation (\ref{eq:fm_pk}) is the
effective volume accessible by the survey at redshift $z$, at
wavenumber $k$ \citep[e.g.][]{tegmark97,sartoris10}.} 

For the matter power spectrum, we adopt the expression for the Cold
Dark Matter provided by \cite{eisenstein98}, which includes the effect
of BAOs. In order to quantify the
information carried by the BAOs detection, we use also the power spectrum
shape from \cite{eisenstein98}, which smoothly interpolates through
the oscillations. Moreover, we study the geometric information carried by the shape of
the power spectrum by describing it  with a general free parameter
$\Gamma$, and thus ignoring its CDM specific relation $\Gamma=\Omega_mh$.
\newline

In our analysis, we assume the following reference values for the
cosmological parameters, consistent with the WMAP-7 best fitting model
\citep{komatsu11}: $\Omega_m=0.28$ for the present-day matter density
parameter, $\Omega_k=0$ for the contribution from the curvature,
$\sigma_8=0.81$ for the normalization of the power spectrum,
$\Omega_{\rm b}=0.046$ for the baryon density parameter, $H_0=70\,{\rm
  km\,s}^{-1} {\rm Mpc}^{-1}$ for the Hubble parameter, $n=0.96$ for
the primordial spectral index. For the DE EoS parametrization of
equation (\ref{eq:eos}), we take $w_0=-0.99$ and $w_a=0$ as reference
values, while for the EDE model of equation (\ref{eq:eos_ede}) we assumed
the reference values of $w_0=-0.93$ and
$\Omega_{e,de}=2\,\times 10^{-4}$. Therefore, we have in total eight cosmological
parameters, which are left free to vary in the computation of the
number counts and power spectrum Fisher Matrices defined in
equations (\ref{eq:fm_nc}) and (\ref{eq:fm_pk}). 
We note that for the above choice of the cosmological parameters, the
reference DE model of equation (\ref{eq:eos}) is consistent with the WMAP-7
results on CMB anisotropies. For the reference EDE model, we
adopt the same values of the non-DE parameters, including
$\sigma_8$. This implies that both models are chosen to have
the same low-redshift normalization as to
provide the same cluster number counts (see also Fig. \ref{fig:rdcs1}), instead of being normalized to CMB.
In the following, constraints on cosmological parameters will be shown
in the $(\Omega_m,\sigma_8)$ and $(w_0,w_a)$ planes, while
marginalizing over the remaining parameters.
\newline

Unless otherwise stated, all the constraints that we present in
the following include the prior information expected from the
measurement of the CMB anisotropies with the Planck
experiment. This prior probability has been computed for each of the two
reference DE models based on equations (\ref{eq:eos}) and
(\ref{eq:eos_ede}).  Cosmological constraints from Planck are derived
by following the description presented in the DETF \cite{albrecht09}
and by using the method described in \cite{rassat08}. We conservatively
assume that only the 143 GHz channel will be used as science
channel. This channel has beam width $\theta_{\rm fwhm}=7.1$ arcmin and
sensitivities $\sigma_T = 2.2 \mu K/K$ and $\sigma_P = 4.2\mu K/K$.
We take $f_{\rm sky} = 0.80$ as the sky fraction in order to reduce
the impact of galactic foregrounds. We use the minimum $\ell$-mode,
$\ell_{\rm min}=30$ in order to avoid problems with polarization
foregrounds. We choose as fiducial parameters $\vec{\theta}=
(\omega_m, \theta_S,\ln A_S, \omega_b, n_S, \tau)$, where $\theta_S$
is the angular size of the sound horizon at last scattering, $\ln A_S$
is the logarithm of the primordial amplitude of scalar perturbations
and $\tau$ is the optical depth due to reionization. After
marginalizing over the optical depth, we calculate the Planck
CMB Fisher matrix in the parameters $(\Omega_m, \Omega_{de},h,
\sigma_8, \Omega_b, {\bf w}, n_S)$ by using the appropriate Jacobian
of the involved parameter transformation \citep{rassat08}. Here $\bf
w$ is a two-component vector which includes the parameters of the two
DE models considered here: ${\bf w}=(w_0,w_a)$ for the model of
equation (\ref{eq:eos}) and ${\bf w}=(w_0,\Omega_{e,de})$ for the model of
equation (\ref{eq:eos_ede}).

\section{Analysis and results}
\label{s:res}

\subsection{Characteristics of the surveys}
\label{s:sur}

In X-ray flux-limited cluster surveys, when the cluster redshift is
known, the flux limit for cluster detection can be translated into a
mass limit, based on a previously calibrated relation between X-ray
luminosity and mass. In order to account for the uncertain knowledge
of this relation, the so-called self-calibration method had been
proposed by different authors \citep[e.g.][
S10]{majumdar03,lima05,cunha09}. In this approach, the uncertainty in
the relation between mass and observable for very large samples can be
described by an intrinsic scatter and a systematic bias in the
estimate of cluster masses. Thus, parameters defining the scaling
relation are treated as fitting parameters to be determined along with
the relevant cosmological parameters. Clearly, a more realistic and
direct method to estimate cluster masses is the one adopted in
\citet{vikhlinin09a} and \citet{mantz10a} using relatively small
samples of clusters based on ROSAT data.  Deep follow-up \chandra\
observations with more than $10^3$ photons for each cluster allowed
them to measure mass proxies which are closely related to cluster
mass, with small ($\mincir 10\%$) intrinsic scatter. Examples of such
proxies are the total gas mass, $M_{gas}$, calculated within a fiducial
aperture radius, or the product of gas mass and temperature,
$Y_X=M_{\rm gas}T_X$, originally introduced by \citet{kravtsov06}.

Next generation of cluster surveys, with high-sensitivity and good
angular resolution such as WFXT, will yield large subsamples of
clusters, each detected with a large enough number of photons to
enable direct measurements of these mass proxies. As additional key
benefit, these high quality data will allow cluster redshifts to be
measured from the Fe-K 6.7 keV line in the X-ray spectra, without
resorting to demanding optical spectroscopic follow-up
campaigns. \citet{yu11} carried out a blind systematic search for
K-shell and L-shell Fe line complex from \chandra\ data. Using a sample
of 46 clusters in the \chandra\ archive, they found that
the cluster redshift can generally be measured from X-ray data with a precision
of $\Delta z\mincir 0.01$ when at least $10^3$ counts are available.


In this paper, we derive cosmological forecasts for three reference WFXT
surveys, which are complementary in terms of sensitivity and sky
coverage: a wide survey covering 20,000 sq.deg. down to a flux
in the [0.5-2] keV band of $f_{\rm 1500}=1.5\times 10^{-13}\fl$, a
medium survey covering an area of 3000 sq.deg. and reaching
$f_{\rm 1500}=3\times 10^{-14}\fl$ and a deep survey over 100
sq.deg. with $f_{\rm 1500}=3\times 10^{-15}\fl$ \referee{(see Table \ref{t:sur})}. \numero
Given the WFXT performances \citep[e.g.][]{giacconi09}, at these
flux limits, clusters are detected with at least 1500 counts, thus
allowing precise measurements of robust X-ray mass proxies and Fe-line based redshifts.

\begin{table} 
\centering
\caption{Characteristics of the envisioned WFXT surveys. Sky coverage
  $\Omega$ (in sq.deg.) (Column 2). Flux limits in the [0.5-2] keV energy band (units of
  $10^{-14}$erg s$^{-1}$cm$^{-2}$) for clusters detected with $\sim30$ counts (Column 3, see also
  \protect\citealt{giacconi09}) and  with  $\geq$\textbf{1500 counts} (Column 4).
Number of clusters, corresponding to $f_{1500}$, detected at $z>0$ 
(Column 5)  and $z>1$ (Column 6).} 
\begin{tabular}{|lccccc|}
\hline
Survey & $\Omega$ & $f_{\rm det}$ & $f_{\rm 1500}$ & $N_{1500}(z>0)$ & $N_{1500}(z>1$)\\ 
\hline 
Wide  & 20000 & 0.5 & 15.0 & 8471 & 0\\
Medium& 3000  & 0.1 & 3.0 & 8435 & 220\\
Deep  & 100   & 0.01 & 0.3 & 1509 &375\\
\hline 
\end{tabular}
\label{t:sur}
\end{table}

To convert cluster fluxes to masses, we follow the same procedure
described in S10 and we refer to that paper for more details. We use
the relation between X-ray luminosity and $M_{500}$ calibrated by
\citet{maughan07}, who analysed \chandra\ data to estimate masses from
$Y_X$ for an extended sample of clusters over the redshift range
$0.1<z<1.3$.


In the following (see Sect. \ref{s:mass}), we show the impact on
cosmological constraints of setting strong priors on the parameters
defining the relation between mass and X-ray observable, when at
least 1500 photons per cluster are available to measure $Y_X$ or
$M_{gas}$ proxies, as well as the cluster redshift.


\subsection{The mass-observable relation}
\label{s:err}

Besides the eight cosmological parameters, our Fisher Matrix analysis 
also
constrains the parameters which specify the redshift dependence of the
fractional mass bias, $B_M$, and of the intrinsic scatter $\sigma_{\ln M}$. 
\referee{Since current data and simulations
  \citep[e.g.][]{stanek_etal10,fabjan_etal11} show no significant
    evidence for a mass dependence of these parameters, in the
    following we do not consider this possibility.}  According to S10,
  we assume:
\begin{eqnarray}
B_M(z) & = & B_{M,0}(1+z)^\alpha \nonumber \\
\sigma_{\ln M}(z) & = & \sigma_{\ln M,0}(1+z)^\beta \,.
\label{eq:nuis}
\end{eqnarray}
In this way, we have four parameters, $B_{M,0}$, $\sigma_{\ln M,0}$,
$\alpha$ and $\beta$, which account for the uncertain knowledge in the
relation between observables and mass (we refer them to hereafter as
\textit{mass parameters}). We consider a reference value
$B_{M,0}=-0.15$ for the mass bias at $z=0$ and $\alpha=0$ for its
evolution\numero. This value of $B_{M,0}$ implies that X-ray masses
are assumed to be underestimated by 15 per cent. This is in line with
the level of violation of hydrostatic equilibrium found by different
authors from the analysis of hydrodynamic simulations of clusters
\citep[e.g.][and references therein]{borgani09}, and from the
comparison between weak-lensing and X-ray masses
\citep[e.g.,][]{mahdavi08,zhang10,okabe10}. We also assume an
intrinsic scatter $\sigma_{\ln M,0}=0.25$, with $\beta=0$ \numero for
its evolution, consistent with the $M_{500}$-$L_X$ relation measured
by \citet{maughan07}. We refer to S10 for a more detailed discussion
on the choice of these parameters. Following \citet{lima05}, we point
out that we use the variance $\sigma_{lnM,0}^2$ and not the scatter as
the varying parameter in our Fisher matrix analysis. In fact,
this variance controls the excess of clusters which are up-scattered
above the selection threshold, with respect to those that are
down-scattered.

In summary, we have four mass parameters that add up to the eight
cosmological parameters for which we compute the Fisher Matrix. 
In order to quantify the effect of the uncertain knowledge of the mass
parameters, we set in the following four different levels of
prior. In order of constraining strength, they can be described as follows.
\begin{itemize}
\item[1.] \textit{No prior}: all the four mass parameters are left
  free to vary by assuming no prior knowledge on their range of variation.
\item[2.] \textit{Weak prior}: we assume $\Delta B_{M,0}=0.05$,
  $\Delta \alpha=1$, $\Delta \sigma^2_{\ln M,0}=0.2$ and
  $\Delta\beta=1$ \numero for the $1\sigma$ uncertainty with which the
  four mass parameters are assumed to be known. The above value of
  $\Delta B_{M,0}$ reflects the current uncertainty between different
  calibrations of violation of hydrostatic equilibrium from
  simulations and from the comparison of weak-lensing and X-ray
  masses. The reference value of $\Delta\alpha$ allows for a large
  variation of the uncertainty with which we can calibrate this
  violation as a function of redshift. The values of $\Delta
  \sigma^2_{\ln M,0}$ and $\Delta\beta$ are rather conservative
  choices, in view of the large number of clusters available from
  future surveys which will allow an accurate estimate of the scatter
  in mass-observable scaling relations. We refer to S10 for a further
  discussion on the choice of this prior for the mass parameters.
\item[3.] \textit{Evolution strong prior}: in order to emphasize the
  role played by the uncertain redshift evolution of the mass
  parameters, we assume in this case the uncertainty in $B_{M,0}$ and
  $\sigma^2_{\ln M,0}$ to be the same as in the \textit{weak prior}
  case, while we assume their evolution to be known to good precision,
  so that $\Delta\alpha=0$ and $\Delta\beta=0$.
\item[4.] \textit{Strong prior}: in this case we consider the
  uncertainties in the calibration of the mass-observable relation are
  so small to be neglected. While this assumption is not realistic in
  an X-ray cluster survey providing only detection of clusters, it is
  plausible for a high-sensitivity survey which provides measurements
  of robust mass proxies for all the clusters above the flux limits
  discussed in the previous section.
\end{itemize}

\begin{table}
\centering
\caption{Prior on mass parameters assumed in the four cases under study 
  (see text) and relative Figure of Merit (FoM) [see equation (\protect
  \ref{eq:fom})] 
  from the combination of three surveys. The analysis is carried out
  by including the Planck prior.}
\begin{tabular}{|lcccc|}
\hline
Reference &&&&\\
values:& $B_{M,0} =-0.15$ &$\alpha = 0$ &  $\sigma_{\ln M,0}^2 = (0.25)^2$ &
$\beta=0$\\ 
\hline
\hline
Cases: & \emph{Strong}  & \emph{Evolution}   &\emph{Weak}  &\emph{No Prior}
\\ 
\hline 
$\Delta B_{M,0}$ &0 &0.05&0.05& / \\
$\Delta \alpha$ &0 &0&1& / \\
$\Delta \sigma_{\ln M,0}^2$ &0&0.2&0.2& / \\
$\Delta \beta$ &0&0&1& / \\
FoM &106&91&64&61\\
\hline
 \end{tabular}
\label{t:nuis}
\end{table}

We summarize in Table (\ref{t:nuis}) these different choices of the
uncertainties in the mass parameters for the different priors that we
assume.
\newline

\begin{figure}
  \includegraphics[width=0.47\textwidth]{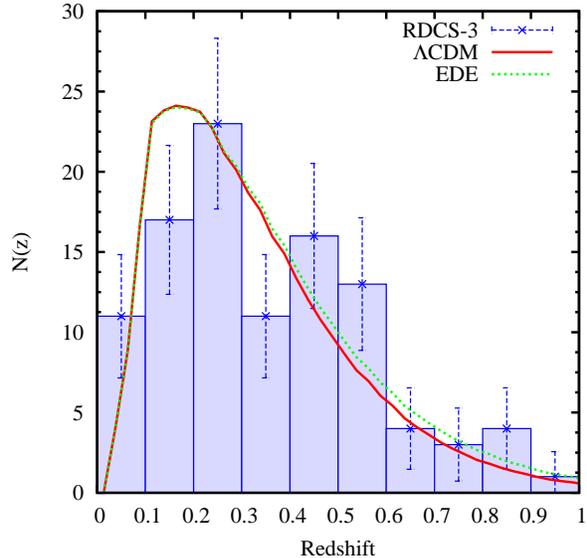}
\caption{Comparison between the observed redshift distribution of
  X-ray selected galaxy clusters (histogram with symbols with
  errorbars) from the ROSAT Deep Cluster Survey -3 
  \protect\citep{rosati02,rosati98} and predictions from the reference DE
  model based on equation (\protect\ref{eq:eos}) (red solid curve) and EDE model
  of equation (\protect\ref{eq:eos_ede}) (green dotted curve).  Errorbars on
  observational data points correspond to $1\sigma$ Poissonian
  uncertainties \protect\citep{gehrels86}.}
\label{fig:rdcs1}
\end{figure}

Before proceeding with the derivation of forecasts for constraints on
cosmological parameters, we verify that our fiducial cosmological
models, with the above reference choice for the mass parameters,
match available observational data on X-ray cluster surveys. To
this purpose, we show in Fig. \ref{fig:rdcs1} a comparison between
the predicted and the observed redshift distribution for the ROSAT
Deep Cluster Survey to an X-ray flux limit of 
$3\times 10^{-14}\fl$ in the [0.5-2] keV band 
(RDCS-3, \citealt{rosati98,rosati02}). We stress
that this is not meant to be a fit to an observational measurement of
the cluster abundance up to $z\sim1$, but rather a test that our
reference model is consistent with current observations. The redshift
distributions for the two reference DE models have been obtained by
convolving the predicted redshift distributions with the
flux-dependent RDCS sky coverage, which provides complete information
on the survey selection function. 
We adopt a minimum luminosity of $L_X {\rm [0.5-2 keV]} = 10^{42}\, \mathrm{erg/s}$ 
which is appropriate for the RDCS selection function.

The good agreement with the data
indicates that the our reference model can be used to provide a
realistic extrapolation of the evolution of the cluster mass function
over redshift and mass ranges which are not probed by currently
available data.

\begin{figure}
\includegraphics[width=0.47\textwidth]{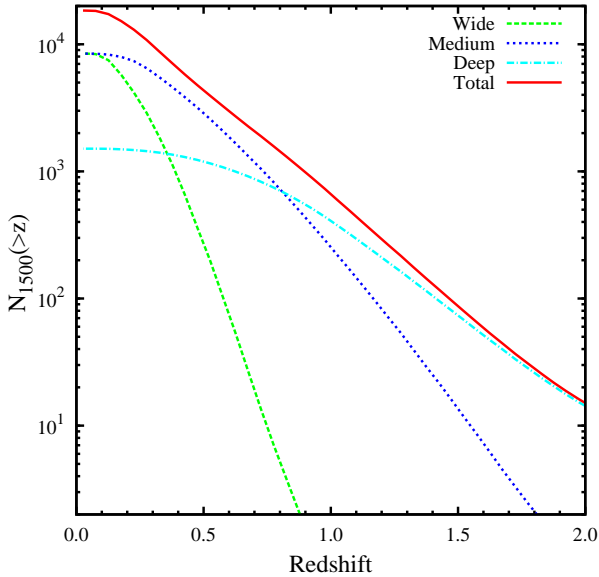}
\caption{The cumulative redshift distributions of clusters detected with
more than 1500 net counts in the three surveys, as predicted by the reference DE model of
  equation (\protect\ref{eq:eos}). Dashed (green), dotted (blue) and
  dot-dashed (cyan) curves are for the Wide, Medium and Deep WFXT surveys,
  respectively, while the solid (red) curve represents the sum of the
  three.}
\label{fig:rd}
\end{figure}
\begin{figure}
\includegraphics[width=0.47\textwidth]{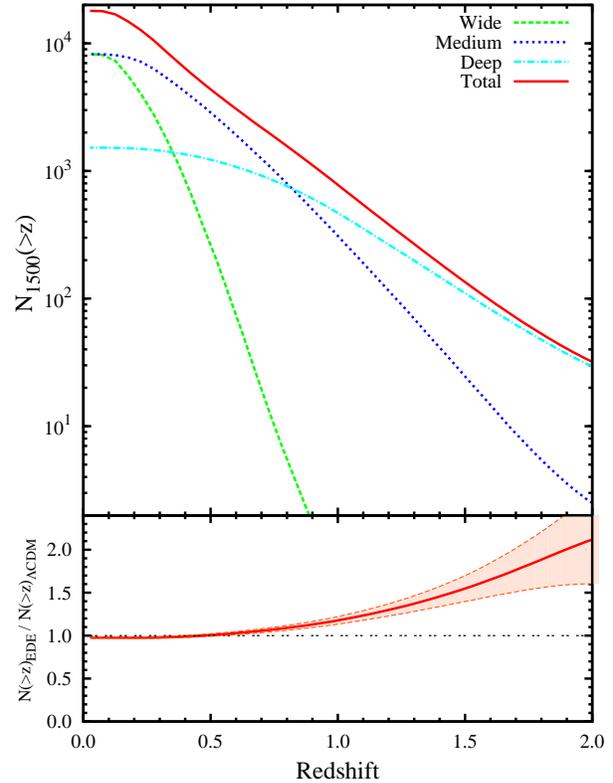}
\caption{The cumulative redshift distributions of clusters detected with
more than 1500 net counts in the three surveys, as predicted by the reference EDE model of
  equation (\protect\ref{eq:eos_ede}). Dashed (green), dotted (blue) and
  dot-dashed (cyan) curves are for the Wide, Medium and Deep WFXT
  surveys, respectively, while the solid (red) curve represents the
  sum of the three. In the bottom panel we show the ratio between the
  cumulative redshifts distributions for the combined survey, as
  predicted by the reference EDE model and the reference DE model
  shown in Fig. \protect\ref{fig:rd}. The shaded region represents
  the Poissonian error on this ratio. }
\label{fig:rdede}
\end{figure}

In Fig. \ref{fig:rd} and \ref{fig:rdede}, we show the
cumulative redshift distributions of clusters to be observed in the
Wide, the Medium and the Deep surveys according to our reference DE and
EDE models, respectively. In the lower panel of Fig.
\ref{fig:rdede}, we also show the ratio between the cumulative
redshift distributions obtained for these two DE models by
combining the three surveys. In order to reproduce the observed
abundance of clusters at low redshift, in EDE models structures start
to form earlier and the halo population follows a slower evolution
than in the \lcdm\ prescription.  We note that the combination of the
three surveys would provide about $2\times 10^4$ clusters with
sufficient photons to allow robust measurements of X-ray mass
proxies. The total number of clusters is dominated by the Wide and the
Medium surveys. The Deep survey is expected to provide a smaller
number of clusters at low redshift, due to the smaller survey
area. However, the number of distant clusters at $z>1$ is dominated by
the Deep survey, owing to its higher sensitivity (see also S10).


\subsection{Cosmological constraints}
\label{s:con}

Having defined the reference cosmological model and the
characteristics of the X-ray cluster surveys, we present in this
section forecasts on constraints of DE EoS parameters. We show
our results in terms of constraints on the $(\Omega_m,\sigma_8)$ and
the $(w_0,w_a)$ plane at the 68 per cent confidence level, after
marginalizing over the other cosmological and mass parameters, and in
terms of Figure of  Merit (FoM). The concept of FoM for DE constraints
was introduced in the Dark Energy Task Force (DETF) report
\citep[e.g.,][]{albrecht09} in order to quantify the knowledge on DE
EoS parameters that future cosmological experiments can reach. In
general, the FoM for the capability of an experiment to constrain a
pair of cosmological parameters $(p_i,p_j)$ can be defined as
\begin{equation}
{\rm FoM}=\left(\det\left[Cov(p_i,p_j)\right]\right)^{-1/2}\,,
\label{eq:fom}
\end{equation}
where $Cov(p_i,p_j)$ is the covariance matrix between the two
interesting parameters. With this definition, the FoM is proportional
to the inverse of the area encompassed by the ellipse representing the 68 per cent
confidence level for model exclusion.

In the computation of the cluster number counts Fisher Matrix,
equation (\ref{eq:fm_nc}), $N_{l,n}$ is calculated within intervals of
observed redshift, with width $\Delta z= 0.05$ out to
$z_{max}=2$. As for the observed mass, we use bins of width $\Delta
\log M= 0.01$, extending from the lowest mass limit determined by the
survey selection function at a given redshift, up to
$10^{16}h^{-1}M_\odot$. We have verified that with this tight binning
in mass we saturate information provided by cluster number counts to
constrain cosmological and mass parameters.

In the computation of the power spectrum Fisher Matrix, given by
equation (\ref{eq:fm_pk}), the average cluster power spectrum defined
by equation (\ref{eq:barpk}) is calculated by integrating over
redshift intervals having constant width $\Delta z=0.2$. \referee{This
  binning, which is coarser than the one adopted for the number
  counts, was chosen as a compromise between the need of extracting
  the maximum amount of information from clustering evolution and the
  request of limiting the covariance between adjacent $z$-intervals
  \citep[e.g.,][]{stril10}}.  Indeed, the contribution from different
$z$-bins can be added in Fisher Matrix defined by equation
(\ref{eq:fm_pk}) only if they carry statistically independent
information. 
\refereebis{Using small redshift bins implies that the
  neighbouring bins are significantly correlated. In this case, the
  covariance terms between different redshift intervals should be included in the
  likelihood function entering in the expression of equation
  (\ref{eq:fm_pk}) for the power spectrum FM.} As for the wave number, we consider a minimum value of $k
= 0.001\,\MpcI$; the choice of this minimum value does not change the
final results, because extremely large wave modes are not sampled by
the surveys used and, therefore, do not provide any contribution to
the Fisher Matrix. The maximum value chosen is $k_{max}=
0.3\,\MpcI$. This choice derives from the need to maximize the
information extracted from the three surveys, while avoiding at the same
time the contribution from small-scale modes where the validity of the
linear bias model is compromised by the onset of non-linearity
\citep[e.g.][see also S10 for a quantitative analysis of the
dependence on $k_{max}$ of FM constraints for non-Gaussian
models]{percival09,stril10,rassat08}.
\refereebis{In particular, \citet{crocce08} studied the non-linear evolution
of BAOs in the Dark Matter power spectrum and correlation function. They showed
that at $k=0.18\,\mathrm{Mpc}^{-1}$ the power spectrum predicted by the linear
theory is lower by a factor of 1.2 at $z=0$  with respect to the
non--linear power spectrum. 
However, we stress that the contribution of information to the Fisher Matrix carried
by the power spectrum at different redshifts 
decreases for both very high and very low
values of $k$ \citep{sartoris10}.
The contribution of the power spectrum directly depends on the 
effective volume,  ($V_{\mathrm{eff}}$ in equation \ref{eq:fm_pk}). 
This quantity depends on the power spectrum itself, which is set by the bias parameter, 
and on the level of Poisson noise, which is set by the number
density of clusters and is maximized at $k \simeq 0.01 \mathrm{Mpc}^{-1}$ (see Fig.7 of S10).}  

Wavenumber bins have been
chosen to have log uniform width $\Delta \log k= 0.1$. Lastly,
introducing redshift space distortions information, the power spectrum
acquires a dependence on $\mu$, which is defined as the cosine of the
angle that \textbf{k} makes with the line of sight (equation
(\ref{eq:kai})). This implies that the Fisher Matrix also involves a
sum on $\mu$ that runs from $\mu=-1$ to $\mu =1$. 
\referee{We choose to divide the interval of $\mu$ into 9 bins. An odd
number of bins is required so that the central bin samples the
values of the redshift space power spectrum computed for $\mu=0$.
In fact, this number of bins maximizes the power spectrum at different
redshifts and wavenumbers and, therefore, the contribution of
$V_{eff}$ (Eq.\ref{eq:fm_pk}) to the Fisher Matrix.  We also
verified that a larger number of bins does not tighten constraints
from the redshift space distortions.}  \newline

\begin{figure*}
\includegraphics[width=0.47\textwidth]{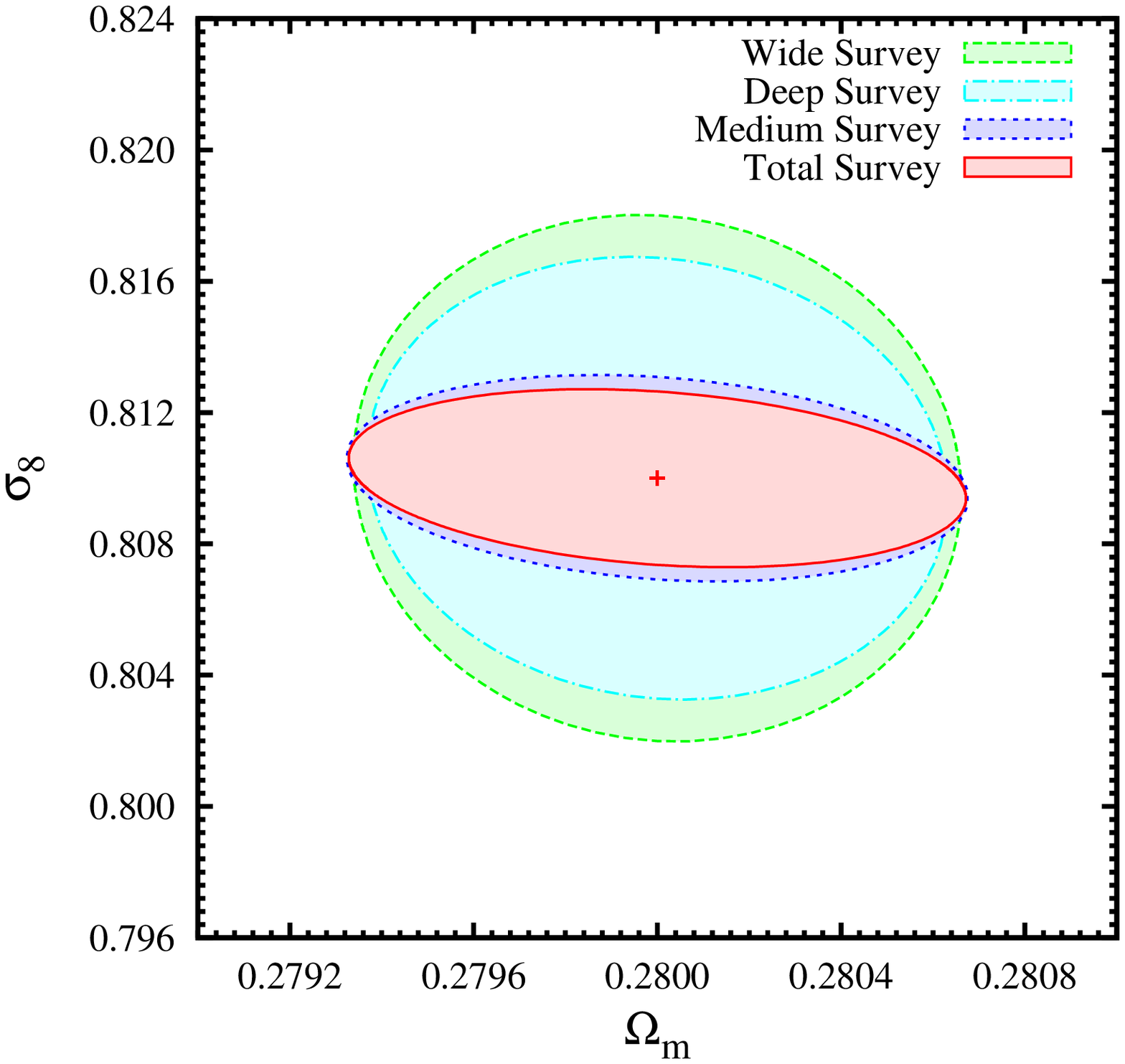}
\includegraphics[width=0.47\textwidth]{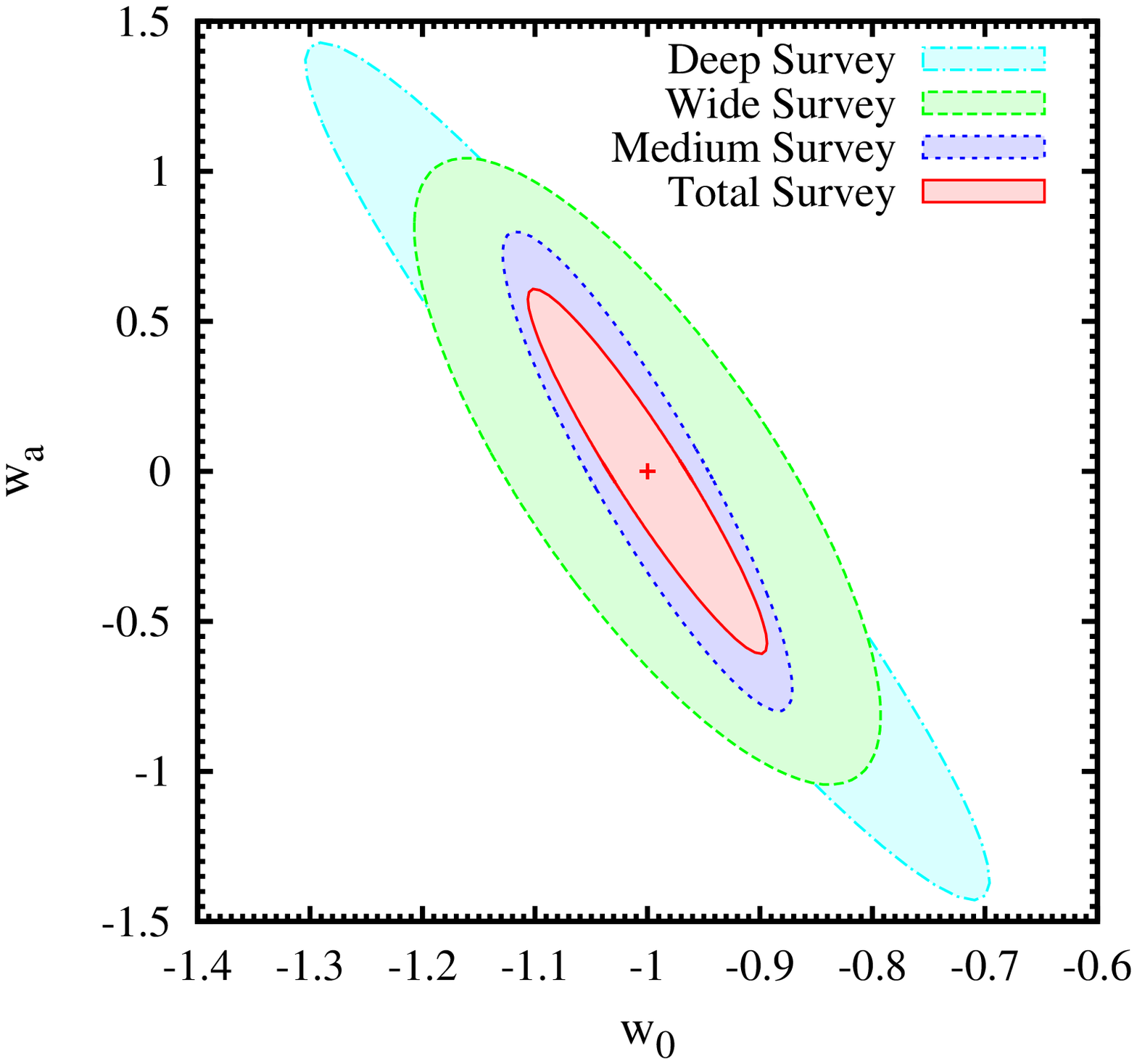}
\caption{Constraints at the 68 per cent confidence level on the
  $(\Omega_m,\sigma_8)$ parameters (left panel) and on the $(w_0,w_a)$
  DE parameters (right panel). In each panel, forecasts for the Deep,
  Medium and Wide WFXT cluster surveys are shown with the cyan, blue and
  green ellipses, respectively, by combining number counts and power
  spectrum information. The red ellipse are the
  constraints obtained from the combination of the three surveys. All
  constraints are obtained by assuming a {\em strong prior} on the
  knowledge of the mass parameters and combining the Fisher Matrices
  for cluster number counts, cluster power spectrum and CMB Planck
  experiment.}
\label{fig:aree}
\end{figure*}

The results of our analysis are presented in Fig. \ref{fig:aree} where we
plot the 68 per cent confidence levels on the $(\Omega_m,\sigma_8)$
and $(w_0,w_a)$ plane, in the left and right panels, respectively. In
each panel, we show the contours obtained for each of the three surveys
and for their combination. Contours are all obtained by combining
information from number counts and power spectrum, also including the
prior information from Planck. A \textit{strong prior} is also assumed
for the knowledge of the mass parameters (see Section \ref{s:err}). 

The results in Fig. \ref{fig:aree} show the trade-off between
surveys area and depth in constraining different cosmological
parameters. As for the results on $(\Omega_m,\sigma_8)$, 
there is no continuous trend in the constraining power of the three
surveys as we reduce the covered area and increase sensitivity. The
Medium Survey is in fact the one with most constraining power,
especially for $\sigma_8$, while the Deep and the Wide Surveys are
somewhat less constraining. Furthermore, the three surveys provide
comparable constraints on $\Omega_m$. This is consistent with the
expectation that constraints on this parameter are mainly provided by
information on the CMB anisotropies, carried by the Planck prior. As
for $\sigma_8$, we remind that this parameter determines the timing of
structure formation, therefore, constraints on its value are sensitive
to both the number of massive clusters included in a survey, and
on the effective redshift range covered by the survey itself. In this
respect, the Medium Survey provides the best compromise between the number
of massive clusters detected within its area and depth.

As for the constraints on $(w_0,w_a)$, we note that their dependence
on the survey area/depth is different from the case of the
$(\Omega_m,\sigma_8)$ parameters. While the Medium survey is still the
most constraining one, we note that the Deep Survey predict a tighter
degeneracy between $w_0$ and $w_a$ than the Wide survey. This
translates into tighter constraints on the redshift evolution of the
DE EoS, if a prior knowledge on $w_0$ is available, consistent with
the fact that the Deep Survey covers a larger redshift interval and provides the larger amount of 
clusters for redshift greater then $0.9$. This
example illustrates that the choice of the survey strategy depends in
principle on the cosmological parameters that one is mostly
interested in. In Table (\ref{t:de}) we report the values of the FoM
and the r.m.s. uncertainty in the DE EoS parameters for each survey
and for their combination, after marginalizing over the other
parameters. The values of the FoM in this table confirm that
Medium Survey alone carries 
most of the contribution to the FoM obtained by combining the three surveys.

\begin{table}
\centering
\caption{Figure of Merit and r.m.s. uncertainty in the DE EoS parameters, 
  for the WFXT three surveys,
  and for their combination. The analysis is carried out
  by including the Planck prior and assuming {\em strong prior} for
  the mass parameters.}
\begin{tabular}{|lcccc|}
\hline
Surveys & Deep & Medium & Wide & Total \\ 
\hline 
FoM & 20 & 60 & 17 & 106 \\ 
$\sigma_{w_0}$ & 0.20 & 0.097 & 0.14 & 0.064 \\
$\sigma_{w_a}$ & 0.94 & 0.54 & 0.70 & 0.41 \\ 
\hline
 \end{tabular}
\label{t:de}
\end{table}

\subsubsection{Effect of mass parameter priors}
\label{s:mass}
\begin{figure*}
\includegraphics[width=0.47\textwidth]{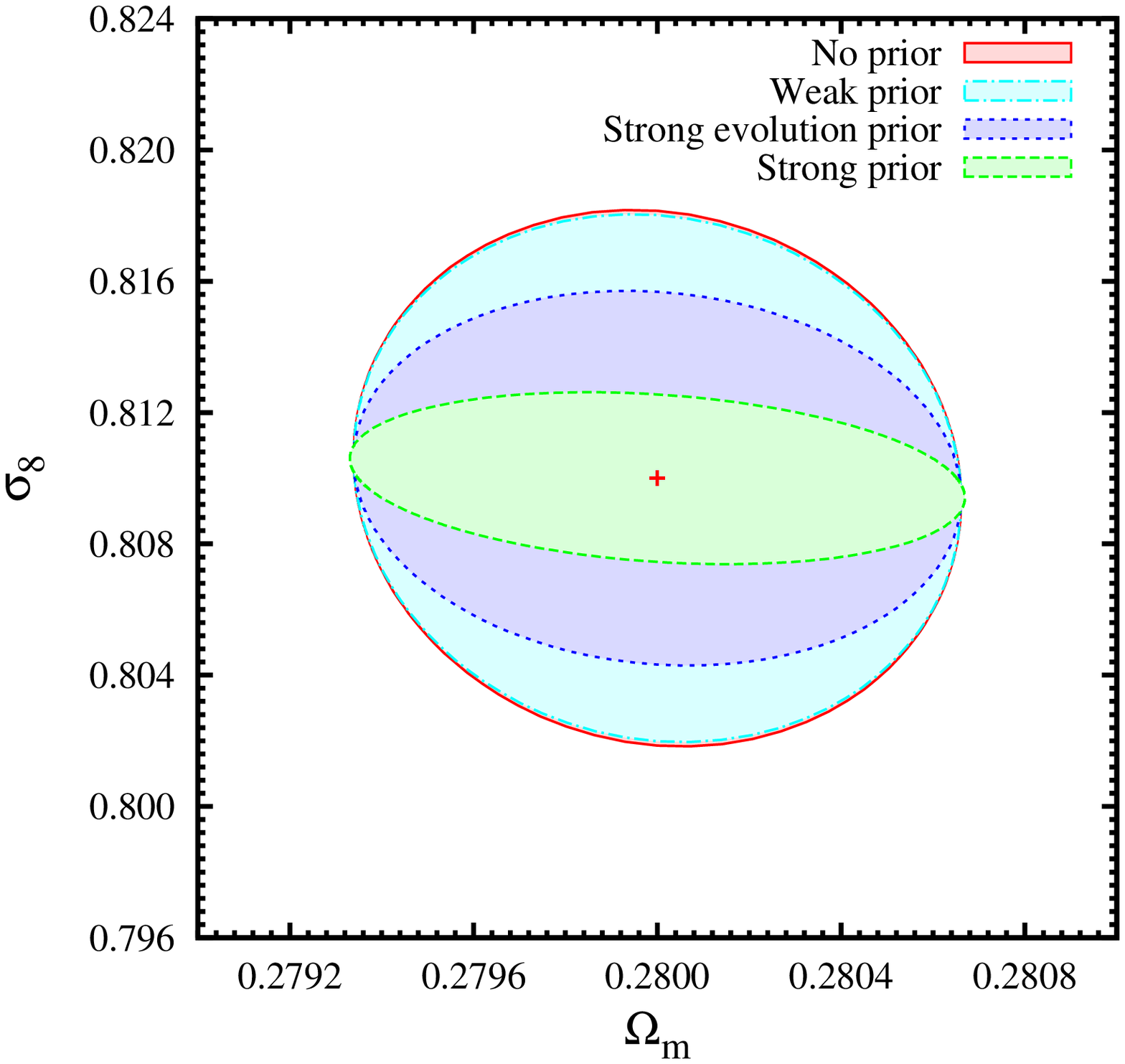}
\includegraphics[width=0.47\textwidth]{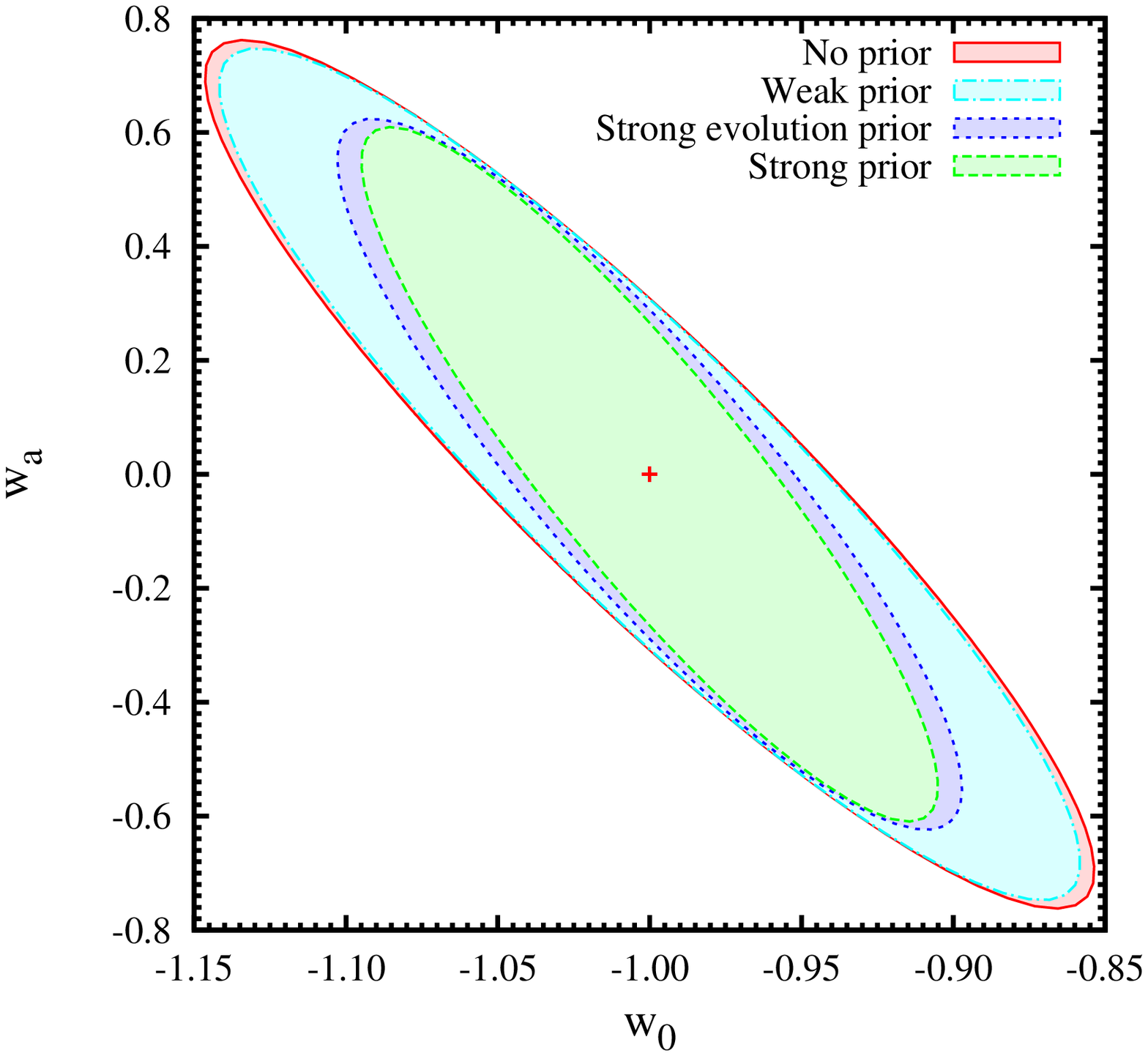}
\caption{Constraints at the 68 per cent confidence level on the
  $(\Omega_m,\sigma_8)$ (left panel) and on the ($w_0$ ,$w_a$)
  (right panel) planes. Different ellipses correspond to different
  assumptions on the prior for the mass parameters (see text): {\em
    no prior} case (red), {\em weak prior} case (cyan),
  \textit{evolution strong prior} case (blue), and {\em strong prior}
  (green). All constraints are obtained by combining cluster number
  counts and power spectrum information for the combination of the
  three surveys.}
\label{fig:nuis}
\end{figure*}

As a first test, we present the effect that using progressively stronger
priors on the mass parameters has on cosmological constraints. The
results of this analysis are shown in Fig. \ref{fig:nuis}, where we
plot the constraints on cosmological parameters obtained by combining
information from cluster number counts and power spectrum, from the
three surveys together. The Fisher Matrix from the cluster experiment
is also combined with the Planck Fisher Matrix.

In the left panel of Fig. \ref{fig:nuis}, we show the constraints
in the $(\Omega_m,\sigma_8)$ plane. A progressively better
knowledge of the relation between X-ray observable and cluster mass
turns into progressively tighter constraints on the $\sigma_8$
parameter, while leaving the results on $\Omega_m$ basically
unchanged. The reason for this behaviour is that constraints on
$\Omega_m$ are mainly determined by the measurement of the CMB
anisotropies and by the shape of power spectrum, which however only
provide rather loose constraints on $\sigma_8$. On the other hand, the
power spectrum normalization is determined by the growth of
cosmic structures, which is traced by the evolution of the halo mass
function. Since a precise measurement of the mass function can only be
obtained through a detailed knowledge of the mass parameters, it is
of little surprise that such parameters determine the accuracy with
which $\sigma_8$ can be measured.

As for the constraints on the DE EoS parameters (see right panel of
Fig. \ref{fig:nuis}), we note that improving the knowledge of the mass
parameters from the {\em no prior} (red ellipse) to the {\em weak
  prior} (cyan ellipse) case only brings a modest enhancement of the
constraining power of the surveys. The main reason for this is that
constraints on DE parameters are here mainly contributed by the
evolution of linear perturbation growth. On the other hand,
constraints on the growth are rather degenerate with the uncertainty
in the redshift evolution of the mass parameters, which is assumed to
be rather generous also in the {\em weak prior} case. \refereebis{Indeed, a more
significant improvement in the constraints on the DE parameters is
obtained for the \textit{evolution strong prior} case (blue ellipse),
which assumes a precise knowledge of the parameters determining the
evolution of the mass-observable relation, with only a slight further improvement in
the constraints obtained for the {\em strong prior} case (green
ellipse). These results confirm the importance of accurately calibrating
  the evolution of the scatter and bias parameters by measuring
  different mass proxies in high redshift clusters selected in the
  Deep survey. Indeed covering at a high sensitivity an even small
sky area allows one to obtain a robust calibration of the scaling
relations between the cluster mass and X-ray mass proxies over a large redshift
baseline. As shown in Fig. \ref{fig:rd}, this survey will provide
about 400 clusters at $z>1$ for which measurements of redshift,
$M_{gas}$ and $Y_X$ will be possible, out of which about 100 are
expected to lie at $z>1.5$.} In Table \ref{t:nuis}, by showing the value of the FoM obtained, we summarize 
the results obtained in our analysis on the effect of uncertainties on the mass parameters in the determination of DE EoS parameters.

\subsubsection{Combining cluster number counts and power spectrum}
\label{s:ncps}

\begin{figure*}
\includegraphics[width=0.47\textwidth]{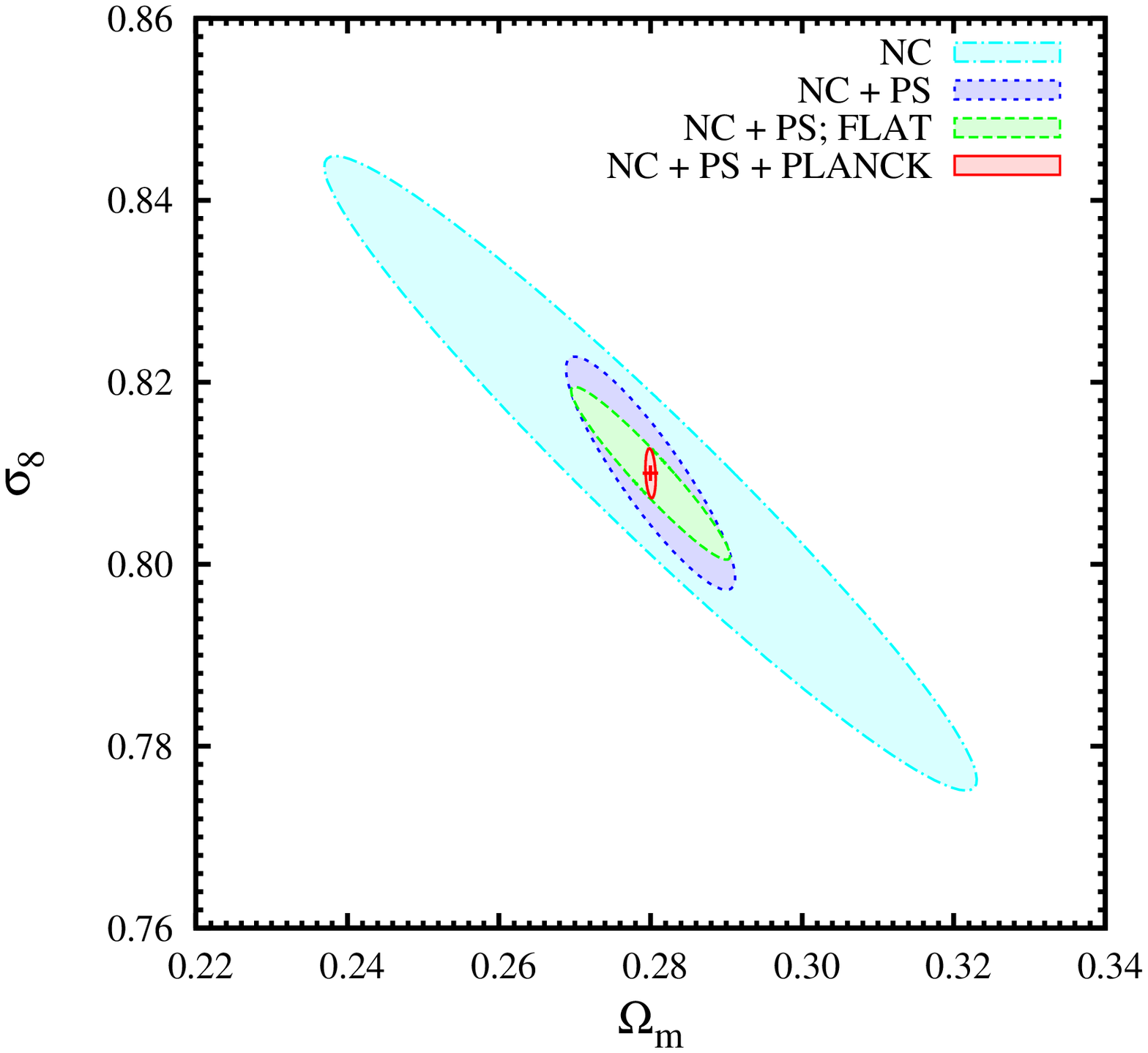}
\includegraphics[width=0.47\textwidth]{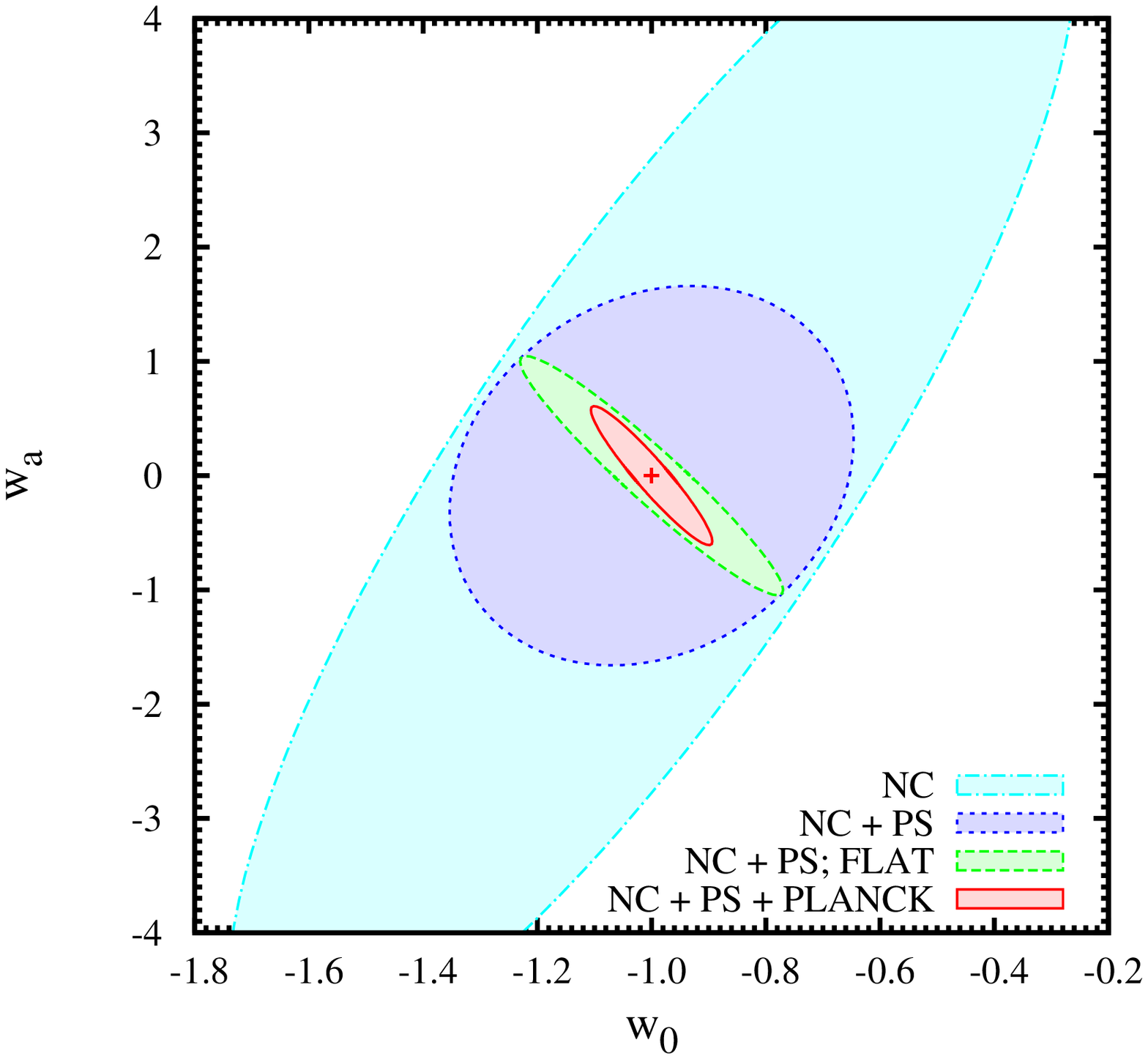}
\caption{Constraints at the 68 per cent confidence level on the
  $(\Omega_m,\sigma_8)$ parameters (left panel) and on the $(w_0,w_a)$
  DE EoS parameters (right panel). Contours, in order of decreasing area,
  are obtained by including the Fisher Matrix from
  cluster number counts only (cyan ellipse), adding cluster power
  spectrum information (blue ellipse), further assuming a flat
  Universe (green ellipse) and adding priors from the Planck
  experiment while leaving geometry free (red ellipse). All
  constraints are obtained by combining information from the three
  surveys and assuming the \textit{strong prior} on the mass
  parameters.}
\label{fig:contr}
\end{figure*}

We discuss now how the combination of number counts and power spectrum
information enhances cosmological constraints. To this aim, we show
the improvement on constraints obtained by adding progressively
information from the cluster number counts, the mean cluster power
spectrum analysis and the CMB prior from Planck. The results are
presented in Fig. \ref{fig:contr} in the ($\Omega_m,\sigma_8$)
(left panel) and $(w_0, w_a)$ planes. Constraints are obtained by
combining information from the three surveys together and assuming
\textit{strong prior} on mass parameters. The redshift evolution of
the cluster number counts sets the direction of degeneracy for the
constraints on \omegam\ and $\sigma_8$. Such constraints are mainly
placed on the linear growth factor of density perturbations through
the mass function. Furthermore, since the density parameters
contributed by matter and DE also affect the expansion history of the
Universe, we expect their values to be constrained by the cluster
number counts, through the redshift evolution of the comoving volume
element. The power spectrum analysis provides information
on the growth rate of cosmic structure through the bias factor, and
the RSDs effect. Moreover BAO features, that depend on the expansion
history of the universe (see Section \ref{s:bao}), and the power spectrum
shape (see Section \ref{s:gamma}) are also sensitive to the underlying DM
distribution.

In Fig. (\ref{fig:contr}), we show that adding the power spectrum information
to the number counts substantially shrinks the
contours. Including the information from the Planck prior (red contour)
further contribute to tighten the contours in the
($\Omega_m,\sigma_8$) plane. In order to verify whether CMB add
information only by constraining the curvature of the Universe, we
also show with the green contour the effect of assuming instead a flat
Universe on the cluster constraints. In this case, results on
($\Omega_m,\sigma_8$) are not drastically improved with respect to the
case in which curvature is a free parameter, while they are
significantly worse than with the Planck prior. The reason for this
result is that CMB anisotropies provide constraints not only on
the curvature, but also on the Hubble parameter $h$, on \omegab\ and on
the primordial spectral index $n_s$. All these parameters enter in
defining the shape of the power spectrum, along with
\omegam. Therefore, their precise determination from the CMB turns into
a significant improvement of constraints on the density parameter
from the shape of the power spectrum.

As for the constraints on $(w_0, w_a)$, their direction of degeneracy
changes as a specific geometry is assumed.
Imposing the flat prior corresponds to fix the redshift at which DE
component, $\Omega_{de}$, starts dominating over $\Omega_m$ and,
therefore, breaks the degeneracy between $w_0$ and $w_a$. By including
the Planck prior, instead of assuming flatness, has a smaller impact
than for the $(\Omega_m,\sigma_8)$ constraints. Therefore, even though
CMB alone does not provide in itself stringent constraints on the DE
EoS, it is quite effective in improving the corresponding constraints
from cluster number counts and power spectrum, owing to its leverage 
on the geometry of the Universe.
\referee{The Deep and Medium surveys dominate the cluster counts at
  $z\magcir 0.4$ (see Fig. \ref{fig:rd}), thus improving the
  constraints on the growth rate of perturbations in a redshift range
  where it is sensitive to the DE EoS.}

\subsubsection{Information from Baryonic Acoustic Oscillations}
\label{s:bao}

We quantify now the geometrical information brought by the presence of
BAOs features in the matter power spectrum. BAOs appear as wiggles
superposed on the power spectrum of the dominant Dark Matter component
\citep[e.g.][]{eisenstein98}. The oscillation scale is proportional to
the inverse of the sound horizon at the matter radiation
equivalence. BAOs carry at lower redshift the same information that
baryonic oscillations in the CMB photon power spectrum provide at the
last-scattering redshift. The position of the wiggles is related to
the amount of dark matter and baryons. As \omegam $h^2$ increases, the
first peak is shifted to higher $k$ values and, moreover, the valleys
and peaks become slightly narrower. The amplitude of the wiggles also
depends on \omegab\ as the oscillations grow stronger as the baryon
fraction increases. In this analysis, we study the constraints on the
$(w_0,w_a)$ DE EoS parameters as obtained by using the transfer
function by \citet{eisenstein98}, which includes BAOs, and by using
instead the transfer function that smoothly interpolates through the
oscillations (see equation (30) of \citealt{eisenstein98}). In the latter case,
the presence of baryons manifests itself only by modifying the overall
shape of the transfer function.

We carried out the analysis including and excluding BAOs in the shape
of the matter power spectrum used to compute cluster number counts and
bias. In order to better appreciate the information carried by BAOs,
in both cases we do not assume any prior on cosmological parameters,
while we use \textit{strong prior} on the mass parameters.
As expected, constraints from Wide survey are those that benefits most
from the presence of the BAOs. This is mainly due to the fact that
this survey provides the best sampling of the long wavelength modes
corresponding to the most prominent first oscillation harmonics. 
The inclusion of the BAOs analysis increases the FoM by
a factor of $2.1$ \numero in this case. On the other hand, no significant
information on BAOs is provided by the Medium and the Deep surveys.
The FoM from the Medium survey does not increase, while the 
FoM provided by Deep survey increases by a factor of 1.2. In
fact, the Deep survey seems to convey slightly more information on the
BAOs. This is mainly due to the higher number density of clusters in this
survey, which reduces the noise when sampling the BAOs.

\subsubsection{Information from the power spectrum shape}
\label{s:gamma}

To quantify the geometrical information encoded in the matter power
spectrum shape, we fit the shape parameter $\Gamma$ in our analysis,
regardless of its dependence on \omegam, $h$ and \omegab\ which is
specific to the type of Dark Matter.
We compare the results obtained by assuming Dark Matter to be Cold with those 
obtained for a general form of Dark Matter. 
In the latter case $\Gamma$ is treated as a free parameter.
In the former, under the CDM assumption, the shape of the transfer function
is given by $\Omega_m h$, that specifies the size of the horizon at
the equality epoch, and by the baryon density parameter. For this
reason, the power spectrum shape carries information on the cosmic
expansion history. Relaxing the CDM assumption, other characteristic
scales could affect the shape of the power spectrum. For
instance, if massive neutrinos provide a contribution to the DM
budget, the power spectrum is expected to be suppressed with respect
to the pure CDM scale on scales smaller than the characteristic
neutrino free streaming scale
\citep[e.g.,][]{hannestad10,marulli11}. Already available
data on the evolution of the cluster mass function have been used to
set interesting constraints on neutrino mass \citep{mantz10}.

\begin{figure}
\includegraphics[width=0.47\textwidth]{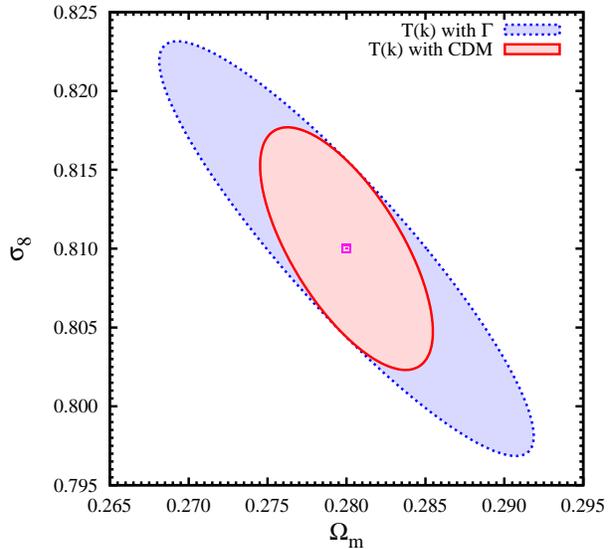}
\caption{Constraints at the 68 per cent confidence level on the
  $(\Omega_m,\sigma_8)$ parameters by leaving the shape parameter
  $\Gamma$ as a free fitting parameter (dotted blue ellipse) or
  assuming its CDM dependence on \omegam, $h$ and \omegab, (solid red
  ellipse). All constraints are obtained by combining information from
  the three surveys together, including number counts and power
  spectrum. In both cases, we assume here the values of
  the parameters $h$, \omegab\ and $n_s$ to be fixed at their reference
  values (see text). No Planck prior are assumed in this case, while
  {\em strong priors} are assumed on the mass parameters.}
\label{fig:bbks}
\end{figure}

In Fig. \ref{fig:bbks}, we show the expected 68 per cent confidence
ellipse on the $(\Omega_m,\sigma_8)$ plane, by combining cluster number
counts and power spectrum information for the three surveys together,
when leaving the shape $\Gamma$ as a free parameter (blue dotted
ellipse) and when using instead its CDM expression (red solid ellipse).
In order to elucidate the effect of a preadopted CDM power
spectrum on these constraints, we fix to their reference values the
parameters that, along with \omegam, determine the power spectrum
shape, namely \omegab, the Hubble parameter $h$ and the primordial
spectral index $n_s$. By removing the assumption of a CDM
spectrum constraints become weaker. In addition, since the effect is more pronounced
for the matter density parameter, the direction of degeneracy 
changes in the sense of a milder dependence of $\sigma_8$ on \omegam.

\refereebis{The shape of the power spectrum is better sampled by the largest survey area covering the widest range of scale. This translates into a stronger constraint on the \omegam\ parameters and consequently on the DE EoS parameters.
Thus, if we relax the assumption of CDM for the shape of the power spectrum, the FoM of
the Wide survey decreases by 7 \numero per cent. This decrement
is less pronounced in the Medium and Deep surveys, whose FoM
decreases respectively by 4 and 5 \numero per cent, owing to their
weaker sensitivity to the spectrum shape.}

\subsubsection{Information from redshift-space distortions}
\label{s:rsd}

In this section, we discuss the effect of including information from
RSDs in the power spectrum analysis. We remind
here that we restrict our analysis to the linear regime, while we do
not attempt to include the non-linear distortions taking place on
small scales. In this case, the dependence of the power spectrum on the angle
between line of sight and wavenumber directions is expressed by
equation (\ref{eq:kai}). The inclusion of the RSDs provides additional
information on the
linear growth rate of perturbations. This test has been amply utilized
to constrain cosmic growth from galaxy redshift surveys
\citep[e.g.,][and references
therein]{guzzo08,wang10,blake11}. However, no evidence has
been reported so far on the detection of such distortions in the
clustering analysis of galaxy clusters. Also, this information has
been never included so far in the derivation of forecasts on the
constraining power of future cluster redshift surveys.

\begin{figure}
\includegraphics[width=0.47\textwidth]{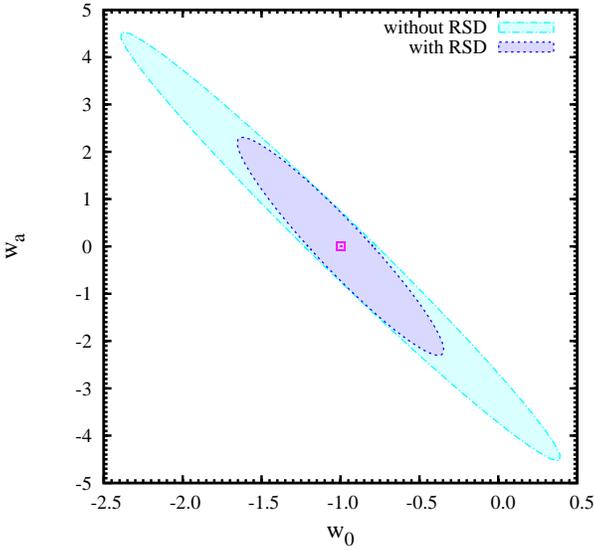}
\caption{Constraints at the 68 per cent confidence level on the
  $(w_0,w_a)$ DE EoS parameters, after including (dotted blue curve)
  or excluding (dot-dashed cyan curve) information from
  redshift space distortions in the cluster power spectrum
  analysis. The constraints are obtained by combining
  information for the three surveys together, including the prior
  information from the CMB Planck experiment, and assuming
  \textit{strong prior} on the mass parameters.}
\label{fig:dis} 
\end{figure}

In Fig. \ref{fig:dis}, we show constraints on the $(w_0,w_a)$
DE EoS parameters obtained by either including (blue dotted ellipse) or
excluding (dot-dashed cyan ellipse) RSDs information in the analysis of
the cluster power spectrum. Both contours represent constraints
derived by combining the power spectrum Fisher Matrix from the
combination of the three surveys, also including Planck prior on
cosmological parameters and \textit{strong prior} on mass parameters.
Quite remarkably, by including information from RSDs 
DE constraints are tightened significantly, thanks to the additional
constraints provided on the linear perturbation growth rate. This leads 
to an increase of the FoM by a factor of $2.2$.
By analysing the three surveys separately, if we do not add Planck prior, we
find that including RSDs
information enhances the value of the FoM by a factor of about $35$, $7.75$
and $6.8$ \numero for the Wide, Medium and Deep surveys, respectively.
The
increasing contribution of RSDs with survey depth is due to the
fact that tighter constraints are obtained by extending the redshift
baseline over which the evolution of perturbation growth is followed.

We emphasize once again that large surveys of galaxy clusters do have
the potential of conveying cosmological information from RSDs. This emphasizes
the importance of obtaining
precise redshift measurements for all clusters included in the survey.

\subsubsection{Constraints on Early DE models}
\label{s:ede}

\begin{figure}
\includegraphics[width=0.47\textwidth]{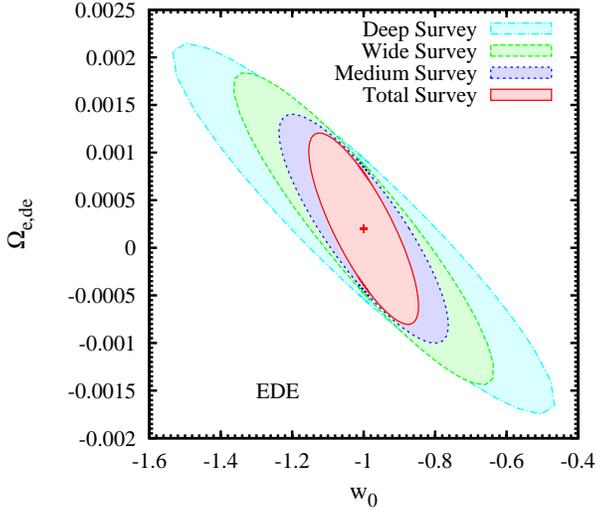}
\caption{Constraints at the 68 per cent confidence level on Early Dark Energy (EDE) EoS
  parameters from the Deep, Medium and Wide surveys (dot-dash cyan,
  dotted blue and dashed green curves, respectively), by combining
  number counts and power spectrum information.  The
  solid red ellipse corresponds to constraints obtained from the combination of
  the three surveys. The Fisher Matrix from Planck experiment and
  \textit{strong prior} on mass parameters are included in the
  calculation of all constraints.}
\label{fig:ede}
\end{figure}

\begin{figure}
\includegraphics[width=0.47\textwidth]{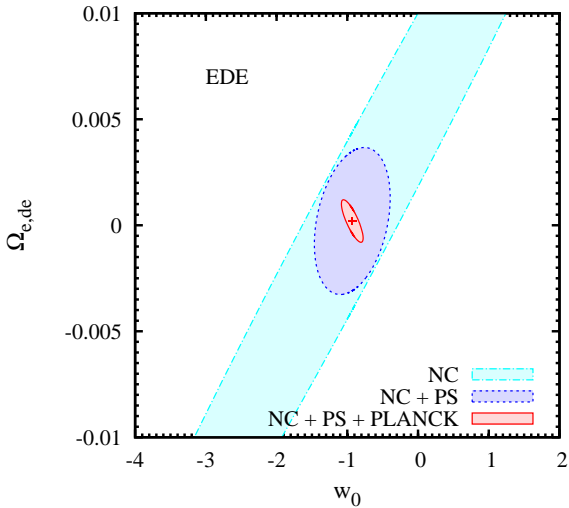}
\caption{Constraints at the 68 per cent confidence level on Early Dark Energy (EDE) EoS
  parameters including the Fisher Matrix from number counts of cluster
  (dot-dashed cyan curve), adding power spectrum (dotted blue curve)
  and adding prior from Planck experiment (solid red curve). All
  constraints are obtained by combining information for the three
  surveys together and assuming \textit{strong prior} on
  mass parameters.}
\label{fig:edecontr}
\end{figure}

As a final analysis, we derive now forecasts for the constraints on
the parameters defining the EoS of Early Dark Energy (EDE) model of
equation (\ref{eq:eos_ede}), which assumes the parametrization by
\citet{grossi09}. In Fig. \ref{fig:ede} and \ref{fig:edecontr},
we show constraints obtained on the ($w_0$, \omegade)
parameters. Cluster number counts and bias are computed by using the
standard mass function by \cite{jenkins01}. As discussed by
\citet{grossi09}, the expression of the mass function calibrated on
N-body simulation according to \lcdm\ model is also a reliable description
of the one provided by simulations of EDE models, at least as
long as DE is homogeneous on small scales \citep[see
also][]{francis09}.

Fig. \ref{fig:ede} presents the constraints obtained for each of
the three surveys and for their combination. They are obtained by
combining cluster number counts and power spectrum information. We include
constraints from the Planck prior and assume \textit{strong
  prior} on mass parameters. \refereebis{The results shown in this figure 
  confirm
that the Medium survey is the one carrying most of the information on
the DE EoS. Thus we can extend to EDE models what we found in the right panel
of Fig. \ref{fig:aree} for models of equation \ref{eq:eos} where DE
influences the cosmic evolution at lower redshift. Even if the Deep
survey should be more sensitive to the EDE thanks to the large number
of clusters at $z > 0.8$, a higher FoM is found for the Medium survey. This is mainly triggered by the number of total clusters provided by the Medium survey even at high redshifts.}

In order to analyse the origin of the constraints on EDE parameters,
we show in Fig. \ref{fig:edecontr} how such constraints
are progressively tightened as we add information from cluster power
spectrum and Planck experiment to the cluster number counts. The latter are
expected to provide rather degenerate constraints on
$(w_0,\Omega_{e,de})$, which is basically associated to the freedom of
choosing a generic geometry of the Universe. The power
spectrum analysis brings in addition both information on geometry through 
the
shape of the transfer function and extra information on
perturbation growth through RSDs. 
\refereebis{It is well known that the primary CMB, with 
the exception of the integrated Sachs-Wolfe effect, does not provide 
constraints on the dark energy for a non-flat Universe \citep{albrecht06,bean01}.
However, the \omegam\ parameter is strongly constrained by the CMB  and this in turn tightens the cluster constrains on DE, as we point out in Section \ref{s:ncps} (see also Figs. \ref{fig:contr}).}
Furthermore, adding 
constraints expected from the Planck experiment causes also EDE constraints to
be much improved, while changing the degeneracy direction.  
\referee{The reason for this is that in the EDE scenario, the purely
  geometrical constraints from CMB anisotropies become slightly more important due to
  a non-negligible DE contribution to the total energy density of the
  Universe at $z\sim10^3$.}  \refereetris{The FoM derived from our constraints increases by a 
factor of 27 when we add information from the Planck experiment to the clusters analysis.} 
In general, this further highlights that
tracing cosmic growth over the widest possible range of redshift is
required in order to tightly constrain the values of DE EoS
parameters \citep{xia09}.

\section{Conclusions}

In this paper, we presented forecasts on the capability of future wide-area
high-sensitivity X-ray surveys of galaxy clusters to yield
constraints on the parameters defining Dark Energy (DE) equation of
states. We considered the standard equation of
state (EoS)
provided by equation (\ref{eq:eos}) and the class of Early DE models of
equation (\ref{eq:eos_ede}). Our analysis was carried out for future X-ray
surveys which have enough sensitivity to provide accurate measurements
of X-ray mass proxies and Fe-line based redshifts for approximately $2\times 10^4$ \numero clusters, thus
extending by more than two \numero orders of magnitude the size of the cluster
samples presently used to derive cosmological constraints
\citep[e.g.,][]{allen11}. We used the Wide Field X-ray Telescope (WFXT)
\citep[e.g.,][]{giacconi09,murray10,rosati10} as a reference
  mission concept along with the Wide (20000 sq.deg.), Medium (3000 sq.deg.) and
Deep (100 sq.deg.) survey
  configurations (see Table \ref{t:sur}).
 We based our analysis on the Fisher Matrix
formalism, by combining information on the cluster number counts and
power spectrum, also including the effect of linear redshift-space
distortions (RSDs). This analysis has been carried out with the main
purpose of dissecting the cosmological information provided by
geometrical and growth tests, which are both included in the analysis
of number counts and clustering of galaxy clusters. The main results
of this study can be summarized as follows.

\begin{figure}
\includegraphics[width=0.47\textwidth]{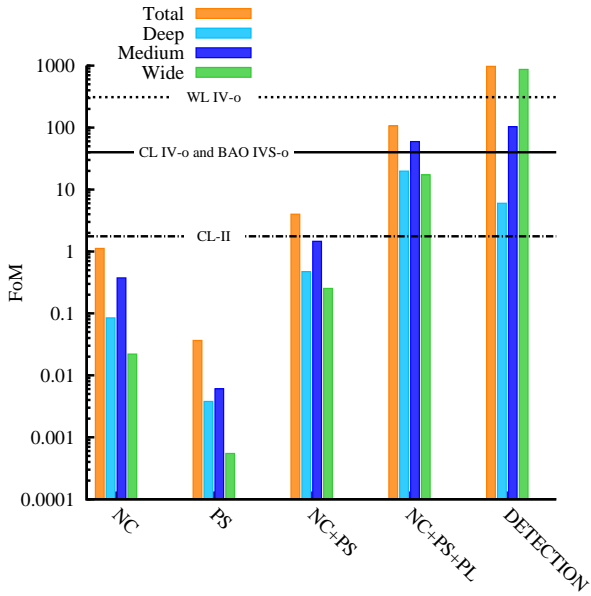}
\caption{The histograms represent the Figures of Merit for the
  $(w_0,w_a)$ parameters from WFXT surveys, as derived in the following configurations:
  by including in the Fisher Matrix the cluster number counts only
  (NC), the cluster mean power spectrum only (PS), the sum of
  the two (NC+PS), and by adding the prior from the Planck experiment
  (NC+PS+PL). All these FoM are obtained by assuming \textit{strong
    prior} on mass parameters. The last group of histograms shows FoMs
  as obtained in the configuration NC+PS+PL by considering all
  clusters that can be detected ($\approx 30$ source counts) in a given survey and by assuming \textit{no prior} on
  mass parameters. The FoM for the Deep the Medium and the Wide
  cluster surveys are shown with the cyan, blue, green histograms respectively. The
  yellow histogram represent the FoM obtained from the combination of
  the three surveys. The horizontal lines show the FoM as reported in
  the DETF \citep{albrecht06} for Stage II Cluster projects (CL-II;
  dot--dashed), for optimistic Stage IV BAO and Cluster projects (BAO
  IVS-o and CL IVS-o, respectively; solid line) and for optimistic
  Stage IV Weak Lensing project (WL IV-o; dotted line), by combining
  each probe with CMB Planck priors.}
\label{fig:fom}
\end{figure}

\begin{description}
\item[(a)] 
  When constraining the parameters of the DE EoS of
  equation (\ref{eq:eos}), we further demonstrate the fundamental
  importance of having a well
  calibrated X-ray observable-mass relation and, most importantly, its
  redshift evolution. 

  We verify that the Figure of Merit (FoM) of the DE EoS increases up to 106 \numero
  when we assume a strong prior on the
  mass parameters, as resulting from a precise and robust calibration
  of the mass-observable relation, with respect to the case in which
  no such prior is available (FoM = 61\numero) (see Table (\ref{t:nuis})).  
  Such an internal calibration can be readily achieved from the same X-ray data 
  by having at least
  $\sim 10^3$ \numero net photon counts for each cluster included in
  the survey.
\item[(b)] We find that the Medium survey is the one carrying most of
  the constraining power (Table (\ref{t:de})), since it is
  expected to yield the largest number of clusters out to redshift
  $z\sim 1$. As such, the Medium survey shows the tightest constraints on the
  evolution of the DE EoS ($\sigma_{w_0}=0.097$ \numero and
  $\sigma_{w_a}=0.54$\numero) and the corresponding highest Figure of
  Merit (FoM $= 60$\numero).  The Deep survey, although covering a
  much smaller area than the Wide survey, adds an important
  contribution to constraining DE parameters (FoM $=20$\numero).
\item[(c)] We quantify the increase of the constraining power from
  the three surveys separately and from their combination, by adding
  progressively information from the cluster number counts, the mean
  cluster power spectrum analysis and the CMB prior from Planck
  experiment. We summarize in Fig. \ref{fig:fom} the resulting
  improvements on FoM. The slightly different
  directions of degeneracy of the constraints in the $(w_0,w_a)$
  parameter space from cluster number counts and power spectrum
  explain why the constraints substantially improve when we consider
  the two contribution together rather then separately. We verified
  that adding the CMB information improve the corresponding
  constraints on the DE EoS, mostly as a consequence of the constraint
  provided by CMB data on the geometry of the Universe (right panel of
  Fig. \ref{fig:contr}).
%

\item[(d)] We find that RSDs carry important cosmological
  information through the linear growth of perturbations, also in the
  case of cluster surveys. Indeed, the DE FoM from the power spectrum
  analysis of the Wide survey increases by a factor 35 when including
  RSDs, while increasing by a factor 7.7 and 6.8 \numero for the Medium
  and the Deep surveys, respectively.
\item[(e)] As for the information carried by the shape of the power
  spectrum, a smaller increase in the FoM is instead measured when
  including BAOs. In this case the FoM from the power spectrum analysis
  of the Wide survey increases by a factor of 2, while no significant
  information on BAOs is provided by the Medium and the Deep surveys.
  Furthermore, relaxing the assumption of CDM and treating the shape
  of the power spectrum as a free parameter reduces the FoM by a
  factor of 1.7 \numero in the analysis of the Wide survey.
\item[(f)] The results obtained for the EDE EoS analysis confirm that
  the Medium survey is the one carrying most of the information on the
  DE EoS. \referee{This emphasizes once more the importance of 
 finding a good balance in the definition of a survey strategy, between the redshift range needed to trace cosmic
  growth and the survey area.}  By extending the
  redshift range of the sample and with the ability to internally
  calibrate the observable-mass relation, we expect to measure the EDE
  EoS parameter (\omegade) with an uncertainty of
  $\sigma_{\Omega_{e,de}} = 6.6 \times 10^{-4}$ \numero.

\end{description}

In order to compare the constraints from the WFXT cluster surveys to
those expected from other cosmological experiments, we also compare in
Fig. \ref{fig:fom} the FoM expected from WFXT to those presented by
\citet{albrecht06} for different large--scale structure probes. In the
DETF report, they showed that Stage II cluster projects (ongoing
surveys) provide FoM $\sim 2$ \numero (Fig. \ref{fig:fom}) when
combined with Planck priors. This analysis was carried out for a
generic cluster count survey covering $200\, \rm{deg^2}$ up to
$z_{max} = 2$, with the simple assumption of a constant mass selection
function. According to Stage IV future experiments, by extending the
survey area to $20000\, \rm{deg^2}$, the FoM rises in the optimistic
configuration \footnote{ As for the clusters analysis, according to
  the optimistic configuration, the mean of the mass-observable
  relation and its variance per redshift interval of $\Delta z = 0.1$
  is assumed in the DETF report to be determined up to a level of
  $1.6\%$.}  up to $\sim 40$ with the contribution of the CMB Planck
priors. We point out that the DETF analysis did not include the
constraints expected from the power spectrum analysis. A similar value
of the FoM was also obtained from an optimistic version of Stage IV
project for BAO analysis based on galaxy redshift surveys, again
including Planck priors. The constraining power of these
optimistic
\footnote{ As for the BAOs analysis, the DETF report introduces the
  $\sigma_F$ parameter, which describes the scatter in the relation
  between the true and the photometric redshifts,
  $\sigma_F^2=Var(z-z_{phot})/(1+z)^2$. In the the optimistic
  configuration $\sigma_F= 0.01$.  } Stage IV experiments is somewhat
weaker than that of the WFXT surveys, with the latter having a FoM
larger by a factor of $\magcir 2$. In
Fig. \ref{fig:fom}, we also show for reference the FoM expected for an optimistic
\footnote{ As for the WL analysis in the DETF report, the r.m.s. bias
  $\sigma_{\ln(1+z)}$ between the mean $z$ and photometric redshift
  for galaxies in $\ln(1+z)$ for each bin of width $0.15$ is assumed
  to be determined with a precision of $0.001$.  Moreover, the shear
  measurement is assumed to be miscalibrated by a factor $(1+f_{cal})$
  that varies independently for each redshift bin. It is assumed that
  the calibration factor of each redshift bin has a Gaussian prior of
  width $\sigma(f_{cal})$. In the optimistic scenario this parameter
  was fixed to $\sigma(f_{cal})=0.001$.}  Stage IV Weak Lensing
experiment, which should reach a value of about 300.

We stress that the above forecasts from the WFXT surveys are obtained
by considering a subsample of clusters with at least 1500 net photon
counts. With this restriction robust mass and Fe-line based redshift measurements
can be readily available from the same survey data, without resorting
to external follow-up calibrations or observations. As such, the
derived constraints should be considered as rather conservative since
they do not include possible information carried by clusters detected
with a smaller number of photons or any other information to constrain
mass from external observations (e.g., Sunyaev-Zeldvich fluxes, weak
lensing masses and optical richness from future surveys). Lowering the
flux limits of the WFXT surveys by a factor of 30 would still
guarantee detection of clusters as extended sources, without however
allowing a measurement of redshifts and robust mass proxies. 
Fig. \ref{fig:fom} shows that by including all the detectable clusters,
the FoM increases by about one order of magnitude, even by assuming no prior on
the mass parameters to compensate for the lack of robust mass
measurements.  
\referee{We note that the Wide survey provides the largest
  constraining power for the DE parameters when we include all
  clusters down to the detection limit. In fact, in this case the Wide
  survey dominates the statistics of clusters counts out to redshift
  1.5 (see Figure 3 in \citet{sartoris10}).}  Clearly, the results
obtained from all the detected clusters must be considered as
optimistic, since they rely on the possibility of confirming all
these extended sources and measuring their redshifts with the aid of
large follow-up observations.

In general, our analysis emphasizes that for large cluster surveys to
be really useful for cosmological applications, not only large samples
are needed but also a robust measurement of mass proxies is required
for a significant fraction of the cluster sample. This will be
possible with future X-ray surveys only with an adequate combination
of survey area, sensitivity and angular resolution. Furthermore, our
results also indicates that the optimization of the survey strategy
depends on the class of cosmological models that one wants to
constrain. For instance, in our previous analysis presented in
\cite{sartoris10}, we showed that the Wide survey is best
suited to constrain deviations from non--Gaussian initial conditions,
due to its ability to sample the long wavelength modes thus detecting
a possible scale-dependence of the bias. This differs from the
conclusion reached in the analysis presented here, where instead we
conclude that the Medium survey is best suited to trace the growth
history of perturbation over a large redshift baseline, as required to
follow the redshift dependence of the DE EoS. As already mentioned in
the introduction, models of modified gravity and of clustered DE
represent an other broad class of models for which clusters can provide
important constraints
\citep[e.g.,][]{rapetti09,Schmidt_etal09,lombriser10}. General predictions of
these models are the scale-dependence of the growth factor of
perturbations and of the bias function
\citep[e.g.,][]{parfrey11}. Large cluster surveys, such as those
considered in this paper, have the potential of placing important
constraints on such signatures of deviations from standard
quintessence variants of $\Lambda$CDM, especially once a suitable
survey strategy is chosen.

The upcoming results from the Planck mission will much improved CMB priors 
based on experimental data. At the same time, the
eROSITA satellite will provide in few years a wealth of fresh X--ray
view of the large scale distribution of galaxy clusters and the
evolution of their population out to $z\sim\! 1$. While waiting for a
first X--ray satellite optimized for surveys, such as WFXT, there is
no doubt that an extensive follow-up campaign will be required
to provide a robust mass calibration with independent methods for a
significant fraction of eROSITA clusters, so as to fully exploit the
cosmological information contained in such surveys.

\section*{Acknowledgements.} We acknowledge useful discussions with
Pierluigi Monaco, Anais Rassat
and with all members of the WFXT team. 
This work has been partially supported by the PRIN-INAF 2009 Grant
``Towards an Italian Network for Computational Cosmology'', by the
European Commission's Framework Programme 7, through the Marie Curie
Initial Training Network CosmoComp (PITN-GA-2009-238356), by the
PRIN/MIUR-2009 grant ``Tracing the growth of structures in the
Universe'' and by the PD51 INFN Grant. BS and SB anckowledge the
hospitality of the Kavli Institute for Theoretical Physics, where part
of this work has been carried out. This research was supported in
part by the National Science Foundation under Grant No. NSF
PHY05-51164. We also acknowledge support from the DFG cluster of
excellence Origin and Structure of the Universe.

\bibliographystyle{mn2e}
\bibliography{dark}

\end{document}